\documentclass[12pt]{article}
\title{Topological Vector Symmetry of BRSTQFT and Construction of
 Maximal Supersymmetry} 
\usepackage{mathrsfs}
\usepackage{amsmath}

\global\arraycolsep=1pt
\usepackage[colorlinks=true,backref=true,linkcolor=black,anchorcolor=black,citecolor=black,filecolor=black,menucolor=black,pagecolor=black,urlcolor=black]{hyperref}

\usepackage[Symbol]{upgreek}
\usepackage{caption}
\usepackage{bbm}
\usepackage{dsfont}
\usepackage{amssymb}
\usepackage{textcomp}
\newcommand{\ba}{/ \hspace{-2.5mm}}
\newcommand{\baa}{/ \hspace{-2.8mm}}

\newcommand{\Scal}[1]{\biggl ({#1} \biggr )}
\newcommand{\scal}[1]{\bigl ({#1} \bigr )}
\def\bea{\begin{eqnarray}}
\def\eea{\end{eqnarray}}
\def\be{\begin{equation}}
\def\ee{\end{equation}}
\newcommand{\CR}{\nonumber \\*}

\newcommand{\gra}[2]{{\scriptscriptstyle (#1 , #2 )}}

\newcommand{\trace}{\hbox {Tr}~}

\def\L{{\cal L}}

\def\Lc{\mathscr{L}}
\def\u{\upsilon}
\def\bu{\bar\upsilon}

\def\a{\mathfrak{a}}
\def\b{\mathfrak{b}}
\def\c{\mathfrak{c}}
\def\d{\mathfrak{d}}
\def\e{\mathfrak{e}}
\def\f{\mathfrak{f}}

\def\bomega{{\overset{\circ}{\omega}}}
\def\bR{{\overset{\circ}{R}}}
\def\bA{{\overset{\circ}{A}}}
\def\bF{{\overset{\circ}{F}}}
\def\bLc{{\overset{\circ}{\mathscr{L}}}}

\def\t{\tilde}

\def\aalpha{{\dot{\alpha}}}
\def\bbeta{{\dot{\beta}}}
\def\deltac{{\delta_{\bar c}}}
\def\usk{{{\frac{1}{|\kappa|}}\,_{\scriptstyle \hspace{-0.5mm} 2}}}

\begin{document}
\allowdisplaybreaks[1]
\renewcommand{\thefootnote}{\fnsymbol{footnote}}

\begin{titlepage}
\begin{flushright}
 LPTHE-05-10\\
 SISSA-24/2005/FM\\
\end{flushright}
\begin{center}
{{\Large \bf Topological Vector Symmetry of BRSTQFT
 and Construction of Maximal Supersymmetry\\}}
\lineskip .75em
\vskip 3em
\normalsize
{\large L. Baulieu\footnote{email address: baulieu@lpthe.jussieu.fr},
 G. Bossard\footnote{email address: bossard@lpthe.jussieu.fr} and 
 A. Tanzini\footnote{email address: tanzini@fm.sissa.it}}\\ 
$^{*\dagger}$ {\it LPTHE, CNRS and Universit\'es Paris VI - Paris VII, Paris,
France}\footnote{
4 place Jussieu, F-75252 Paris Cedex 05, France.} 
\\
$^{* }$\it Department of Physics and Astronomy, Rutgers University \footnote{
Piscataway, NJ08855-0849,~USA.} 
\\
$^{\dagger\dagger }${\it SISSA, Trieste, Italy}
\footnote{via Beirut 2/4, 34014ÐTrieste, Italy} 
\\
\vskip 1 em
\end{center}
\vskip 1 em
\begin{abstract}

The scalar and vector topological 
Yang--Mills symmetries determine a
closed and consistent sector of Yang--Mills supersymmetry. 
We provide a geometrical construction of these symmetries, 
based on a horizontality condition on reducible manifolds.
This yields globally well-defined scalar and vector
topological BRST operators. These operators 
generate a subalgebra of
maximally supersymmetric Yang--Mills theory, which is small enough to be closed
off-shell with a finite set of auxiliary fields and 
 large enough to determine the Yang--Mills supersymmetric theory. 
Poincar\'e supersymmetry is reached in the limit of flat manifolds. 
The arbitrariness of the gauge functions in BRSTQFTs is thus removed by the 
requirement of scalar and vector topological 
symmetry, which also determines the complete supersymmetry
transformations in a twisted way. 
Provided additional Killing vectors exist on the manifold, an
equivariant extension of
our geometrical framework is provided, and the resulting ``equivariant
topological field theory'' corresponds to the twist of super
Yang--Mills theory on $\Omega$ backgrounds.

\end{abstract}

\end{titlepage}
\renewcommand{\thefootnote}{\arabic{footnote}}
\setcounter{footnote}{0}



\renewcommand{\thefootnote}{\arabic{footnote}}
\setcounter{footnote}{0}



\section{Introduction}

Topological Yang--Mills theories have been studied extensively in
various dimensions some years ago \cite{BS,donald,et,enfin}. 
 They can be
 defined as a BRST invariant gauge-fixing of a
topological invariant, and their topological observables are
determined from the cohomology of the topological BRST scalar
symmetry, whose geometrical interpretation is well understood.  

However, a yet unsolved mystery is their relation, by a twist operation, to
Poincar\'e supersymmetric theories, which describes particles. There is good evidence that this
relation also extends to the case of topological gravity versus
supergravity \cite{sugra}. In fact, since topological symmetry has a 
clear geometrical interpretation, it has been proposed to use it to {\it
 define} Poincar\'e supersymmetry. Here we reach an 
understanding of the so-called vectorial topological symmetry of TQFT's, which
further support this idea.

 Vector 
symmetry was first observed as an invariance of the Chern--Simons
action, gauge--fixed in the Landau gauge \cite{thomson}. Its existence
can be heuristically guessed from the possible conservation of the
BRST 
antecedent of the energy momentum tensor. For a topological action
that is the twist of a supersymmetric theory, its expression is
identical to the symmetry that one obtains by 
 twisting the spinorial generators of Poincar\'e supersymmetry, (as
 for the case of the scalar topological BRST operator). In fact, the
 twisted formulation has been used to greatly improve the study of 
various non--renormalization properties
of $\mathcal{N}=2$ and $4$ supersymmetric theories \cite{Tet}. 

This paper focuses to the Yang--Mills case. We show that the 
vectorial topological symmetry can be directly introduced,
geometrically, prior to the construction of the TQFT. Basically, the
vector symmetry arises when one associates reparametrization symmetry
and topological 
symmetry in a relevant way. It is important to work on manifolds that
contain at least one covariantly constant vector. Eventually, the
superYang--Mills theory, with 
 Poincar\'e supersymmetry, is reached by untwisting the theory, 
in the limit of flat
 manifolds. 

We also use the method for constructing 
``equivariant topological field theories'', whose observables
are related to the equivariant cohomology classes of the moduli space
of instantons. In fact, these topological theories can be seen as the twisted
versions of the Super Yang--Mills theories on the $\Omega$
background introduced in \cite{nikita}, that are deformed version of
ordinary supersymmetric theories.
 
The scalar and vector invariances of TQFT's constitute
a relevant subalgebra that can be closed ``off-shell''.
 Eventually, this subalgebra is sufficient to completely 
determine
the full ``on-shell'' set of supersymmetry generators 
in the flat space limit.
We actually show that the invariance under scalar and vector
symmetry, which we will geometrically construct, is sufficient in order to fully determine the $\mathcal{N}=2$
supersymmetric action, in 8 and 4 dimensions, respectively. In the
later case the supersymmetry with its 8 generators is actually
determined by the construction of 5 generators, which build a closed
algebra, and in the former case, 9 generators are sufficient to
determine the supersymmetry with its 16 generators. The rest of the
generators, (they are self-dual tensor in the twisted form), can be
considered as an effective symmetry, that one gets for free as an
additional symmetry of the action. They complete the scalar and vector
symmetry generators into a set that can be untwisted toward Poincar\'e
supersymmetry. They have no geometrical interpretation in our
knowledge, and, moreover, the complete set of generators cannot be
closed off-shell for the case of maximal supersymmetry, with 16
generators. 

 Determining the TQFT, and afterward the supersymmetric theory, from a
 symmetry principle that has a clear geometrical meaning, appears to
 us as a progress. 
 Indeed, in earlier works, after having the geometrical
construction of the scalar topological BRST symmetry, the
determination of the action was tantamount to that of ``topological
gauge functions'', including self--duality equations, but was not
relying on a symmetry 
principle. Rather, one was looking for self--duality equations for the
gauge fields, which one can enforce in a BRST invariant way. 
The lack of a complete symmetry principle was
frustrating in this construction. 

The way the geometrical
construction for the scalar and vector BRST symmetries works is
through the construction 
of two nilpotent graded differential operators $s$ and $\delta$ that
are nilpotent and anticommute up to a Lie derivative. These
operators have a transparent meaning in the fiber bundles where the 
Yang--Mills field and the classical gravitational field are naturally
defined. At each step of our construction,
the necessary requirement of global well definition can be
checked. 

There is eventually a duality symmetry between 
$s$ and $\delta$, which merely express a symmetry between topological
ghosts and antighosts. This gives a better understanding of antighosts
as geometric entities, instead as BRST antecedents of the Lagrange
multipliers of gauge functions, as has been traditionally done,
through the notion of trivial BRST quartets. 

In the generic case of 8 dimensions for maximal supersymmetry, the
necessity of having a
manifold with a constant vector implies that its holonomy group be 
$G_2$, or a subgroup of $G_2$. 

The formula that we will obtain are very similar in four and eight
dimensions. Our results can be extended
in 
lower dimensions by using dimensional reduction. Besides $G_2$-manifolds, 
cases of interest are Calabi-Yau manifolds, provided they contain a covariantly constant vector.
Eventually, we express the topological actions as a
$s \delta$-exact (i.e, scalar {\it and } vector BRST-exact) term, with
a nice correspondence with the Chern--Simon action. In fact our formula are reminiscent of previous one found in the general context case of ``balanced topological theories", \cite{moore}, with $\mathcal{N}_T \geq 2$, but here we manage to consider the case $\mathcal{N}_T =1$, with a pair of ``balanced operators" ($s$ and $\delta$), by using a covariantly constant vector in the manifold. Then, it is quite natural to write supersymmetric actions as a $s \delta$-exact term, but the potential cannot be generally considered as a Morse function, and it allows for ghosts and antighost that different tensor structures.

The paper is organized as follows. We first give heuristic evidence 
that the vector symmetry is a consequence of the possible conservation of the
BRST-antecedent of the energy-momentum tensor of a TQFT - in fact it is equivalent. Then we give the important result that there is a
geometrical construction of the vector symmetry which is completely independent of the idea of Poincar\'e supersymmetry. (We include a section
for ensuring global consistency of the formula). We display the
invariant action under various forms, and briefly discuss the untwisting toward supersymmetry.
In the last part of the paper, we show the 
equivariant extension of our formulation and show the relationship
with twisted supersymmetry on the $\Omega$-background.


\section{ Physical evidence for the existence of a vector symmetry
 $\delta$ in a BRSTQFT} 
\label{VS}

Let $S $ be a BRSTQFT topological action. Its Lagrangian is
reparametrization invariant. In local coordinates, ${\L}_\xi$
represents the 
action of diffeomorphisms on the fields, and we can define local
functionals $L_{A\,\mu} (x-y)$ corresponding to each field
$\varphi_A$ of the theory as follows:
 \be
 { \L}_\xi \varphi_A (x) = -\int_M d^n y 
\xi^\mu(y) L_{A\, \mu} (y-x) 
\ee
 We are aware that in order this operator be globally well-defined, we should add
to the Lie derivative $\L_{\xi}$ gauge transformations that permit its 
 covariantization. However, in this section, we only consider 
operators that are basically defined modulo gauge transformations, in such a way that this
subtlety is not relevant. Global requirements will be fulfilled when we will construct the vector symmetry, the existence of which we heuristically justify in this section. The reparametrization invariance of $S$ implies the ``off-shell" conservation law :
 \be \nabla^\nu T_{\mu\nu}(x) = \frac{1}{\sqrt{g}} \sum_A \int_M d^n y \, 
L_{A\,\mu} (x-y) \frac{\delta^L S}{\delta
 \varphi_A (y)} \ee
where $\delta^L$ denotes the left--derivative, and
$T_{\mu\nu} $ is the energy momentum tensor
 \be T_{\mu\nu} = \frac{1}{e} \,\delta_{ab}\, 
 \frac{\delta S}{\delta e^\mu_b} \,e^a_\nu\label{EMT}\ee
Up to a topological term, the action $S$ is $s$-exact,
$S=s \int_M \Uppsi$, where the topological gauge function 
$\Uppsi$ has ghost number $-1$ and $s$ is a topological BRST
operator, which is a scalar under reparametrization. The $s$-exactness of the action implies that the energy momentum tensor is also $s$-exact. Thus :
\be\label{cons} T_{\mu\nu} = s \Lambda_{\mu\nu} \ee
where $\Lambda_{\mu\nu}$ is a local functional of the fields and has
ghost number -1. The gauge function $\Uppsi$ is yet arbitrary. Our aim is of
removing this indetermination by a symmetry principle. The later should also be a canonical
property that defines a regularized action in the path integral. 

We propose that this additional requirement is that the energy
momentum tensor admits a conserved BRST antecedent, modulo
equations of motions, consistently with Eq.(\ref{cons}): 
\be\label{prop} \nabla^\nu \Lambda_{\mu\nu} \approx 0 \ee

As a matter of fact, this property (\ref{prop}) determines the theory in the 
Yang--Mills case, by adjusting all coefficients in the possible topological gauge functions $\Uppsi$, in such a way that one eventually gets the twisted Yang--Mills supersymmetric action.

Eventually, we will transform this property into a symmetry principle.

 Since $\Lambda_{\mu\nu}$ is a local functional of the fields, its
conservation law must take the form 
\be \nabla^\nu \Lambda_{\mu\nu} (x) = \frac{1}{\sqrt{g}} \sum_A \int_M
d^n y \, V_{A\,\mu} (x-y) \frac{\delta^L S}{\delta \varphi_A (y)}
\label{conL}\ee 
where $V_{A\,\mu} (x-y)$ are local functionals with ghost number $-1$ of 
the fields $\varphi_A$. As in the case of
the energy momentum tensor, where $ L_{A\,\mu}$ determines the
diffeomorphism generators, we can define from the $ V_{A\,\mu}$ the
following vector operator \cite{vSusyP, vSusy} : 

\be\label{defd}
\delta \varphi_A (x) = -\int_M d^n y \, 
\kappa^\mu(y) V_{A\, \mu} (y-x) 
\ee
Here $\kappa^\mu(y)$ is a globally well defined {\it given} vector
field (which makes a distinction as compared to the ghost $\xi^\mu(x)$ of the diffeomorphism
symmetry). Note also that this transformation is not an infinitesimal
one, $\kappa$ is a finite vector field. Asking for 
the invariance of the action $S$ under $\delta$-transformations
restricts the choice of $\kappa$. One has indeed: 
\bea
\label{mah}
\delta S &=& -\int_M d^n x \,\sum_A \int_M d^n y \, 
 \kappa^\mu (y) V_{A\,\mu} (y-x)
 \frac{\delta^L S}{\delta \varphi_A (x)}
= \int_M d^n x\, \sqrt{g} \, \Lambda_{\mu\nu}
\nabla^{\nu} \kappa^\mu \CR
\eea
(In the last equality we performed an integration by parts, so the necessity of global consistency must be remembered).
From Eq.~(\ref{mah}) we deduce that the $\delta$-invariance of the 
action only holds if
$\kappa$ is covariantly constant.
We stress that, 
$\kappa$ being a Killing vector
is sufficient, since $\Lambda_{\mu\nu}$ is generally non
symmetrical, see Sect.5 for more details. 

The association between a global symmetry
of the theory and the conservation of a current is nothing but the 
 Noether theorem. As a matter of fact, on any given specific case,
 one can redefine 
$\Lambda_{\mu\nu}$ by the addition of a term linear in the equations of
motion, in such way that its conservation law takes the simpler
form
\be
\nabla^\nu \Lambda_{\mu\nu}'(x) = \sum_A V_{A\,
 \mu}(x)\, \frac{\delta^L S}{\delta \varphi_A (x)} 
\ee
Under this form, $\Lambda_{\mu\nu}'$ can be identified to the Noether
current associated to the $\delta$ symmetry.

 The statement that the
conservation equation determines the complete form of the operator
$\delta$ is however a non trivial one. As we will see further down, in this heuristic derivation, 
$\delta$ is determined modulo gauge transformations and terms linear
in the equations of motion. 

The understanding of the vector symmetry requires the determination of 
its commutation relations with the scalar BRST symmetry. We have :
\bea 
\label{id}
\int d^n y L_{ \mu A} (x-y) \frac {\delta S}{\delta \varphi_{A } (y)} &=& \sqrt{g}
\, \nabla^\nu T_{\mu\nu} (x) =
 s \int d^n y V_{{ \mu A } }{ (x-y) } \frac{\delta 
 S}{\delta \varphi_{A } (y)} \CR 
&=& \int d^n y \Scal{ s V_{{ \mu A}}{(x-y) } \, \frac{\delta 
 S}{\delta \varphi_{A } (y)} - (-1)^A V_{{ \mu A }}{(x-y) } s \frac{\delta 
 S}{\delta \varphi_{A }(y) }} \CR 
&=&\int d^n y \Scal{ s V_{\mu A }(x-y) + \int d^n z V_{\mu B }(y- z)\frac{\delta 
 s \varphi_{A }{( y)} }{\delta \varphi_{B }{( z)} }} \, \frac{\delta 
 S}{\delta \varphi_{A }{( y)} } 
 \CR
\eea
where the sum over repeated indices is assumed, as well as the fact
that the functionals derivatives are taken to the left.
To prove Eq.~(\ref{id}), one uses the following identity, which is consequence of the $s$-invariance of $S$:
\bea
0 &=& \int_M d^n z \frac
{\delta } 
{ \varphi_{A } (y) }
\Scal{
s\varphi_{B } (z)
{ \frac{\delta S}{\delta \varphi_{B } (z)}}
}\CR
&=& \int_M d^n z \Biggl(
 \frac
{\delta s\varphi_{B } (z) }
{\delta \varphi_{A } (y) }
{ \frac{\delta S}{\delta \varphi_{B } (z)}}
+ (-1)^{(A+1)B} s\varphi_{B } (z)
\frac{\delta } 
{ \varphi_{A } (y) }
{ \frac{\delta S}{\delta \varphi_{B } (z)}} \Biggr)
\CR
&=& \int_M d^n z \frac
{\delta s\varphi_{B } (z) }
{\delta \varphi_{A } (y) }
{ \frac{\delta S}{\delta \varphi_{B } (z)}}
+ (-1)^A s
{ \frac{\delta S}{\delta \varphi_{A } (y)}}
\eea
In fact, Eq.~(\ref{id}) indicates that one has 
\be\label{com}
\{ s, \delta \} \approx \L_\kappa
\ee
on all fields $\varphi_{A}$, at least modulo gauge transformations
and modulo terms proportional to the equations of motion. 

This point is not completely obvious. 
Eq.~ (\ref{com}) would be exact if the
Eq.~(\ref{id})\, would constrain the both functionals 
$L_{\mu\, A} (x-y)$\, 
and \,\,$s V_{\mu A }(x-y)+\int d^n z V_{\mu B}(y-z)\frac{\delta s
 \varphi_{A}{(y)}}{\delta\varphi_{B}{(z)}}$ to be equal. The
indetermination of this equation reverts to the determination of the
solution of the equation 
\be
\int d^n y R_{\mu\, A} (x-y) \frac{\delta S}{\delta \varphi_A(y)} =
0 \label{Req} 
\ee 
and the equation (\ref{com}) is satisfied modulo the transformations
which could be generated by the functional $R_{\mu\, A}$. One has actually an analogous situation for the determination of the
operator $\delta$ from the conservation low of $\Lambda_{\mu\nu}$.

We first observe the existence of the following solution of Eq.~(\ref{Req})
\be
R_{\mu\, A} (x-y) = \Scal{ R_{\mu\, AB} (x-y) - (-1)^{AB} R_{\mu\, BA} (x-y) }
\frac{\delta S}{\delta \varphi_B(y)} 
\ee 
for any local functional of the fields $R_{\mu\, AB}$. Because of this
solution the commutation relation of $s$ and $\delta$ could be true only
modulo terms involving the equations of motion. One has to check whether there are 
 other solutions of the equation (\ref{Req}) that cannot be written
as a term linear in the equations of motion. We assume that all local
functionals which are zero when the equations of motion are satisfied
are linear in the equations of motion thereself via a local functional
of the fields. With this assumption, if there is another solution, we
can differentiate the equation (\ref{Req}) with respect to
$\varphi_B(z)$, and obtain
\be \Biggl( \int d^n y R_{\mu\, A} (x-y) \frac{\delta^2 S}{\delta
\varphi_A(y) \delta\varphi_B(z)} \Biggr)_{\left| \frac{\delta
S}{\delta \varphi_A} = 0 \right .} = 0 \label{gd}
\ee
when the equations of motion are satisfied. The functional
$\frac{\delta^2 S}{\delta \varphi_A \delta\varphi_B}$ is only degenerated in
theories with constraints. The solution of the equation (\ref{gd}) is
by definition a gauge transformation. So the general solution
of the equation (\ref{Req}) is a sum of terms linear in the equations of
motion, and of local functionals which correspond to gauge
transformations (or reparametrizations in gravity). Therefore $\delta$
and its commutator with $s$ are determined modulo gauge
transformations and equations of motion from (\ref{conL}) and (\ref{id}).

At this level of the discussion, one may feel frustrated by the lack
of geometrical understanding of the situation, and it appears that a
direct construction of $\delta $, which satisfies Eq.(\ref{com}) and
has ghost number -1, is needed.

Therefore we now adopt the attitude that one must reverse the point of view, and directly construct both differential operators 
$s$ and $\delta$, from geometrical principles. Then the
determination of the action from $s$ and $\delta$ invariances will warranty
 the conservation law of both the energy momentum tensor and of its $s-$antecedent 
$\Lambda_{\mu\nu}$ \footnote{The
 question of anomalies of $s$ and $\delta$ invariances is of course
 an interesting question}. The determination of the superPoincar\'e algebra will be a corollary, using twist arguments that are allowed on the manifold that we will use.

The following sections are devoted to the geometrical construction
of the symmetries in the
Yang--Mills case, for the generic dimensions 4 and 8. We will also construct ab-initio the differential operators $s$ and $\delta$, 
with an interesting irruption of antighosts on the scenery. In fact, their geometrically interpretation will arise form a duality relation between the ghosts and the antighosts, as in balanced topological field theories.

 The algebra will respect by construction the closure relation Eq.(\ref{com}) 
 that is suggested by 
 the above heuristic discussion. Eventually, we will compute the
 antecedent of the 
energy momentum tensor, and verify that it generates the $\delta$-symmetry. 
\section{The gravitational and Yang--Mills horizontality condition
 for the scalar and vector topological symmetries} 
\label{Hcon}
\subsection{ Topological symmetry and globality requirement}
 
The Yang--Mills topological symmetry BRST operator $s_{\rm top} $ is defined from the equation 
\be\label{basi}
( d + s_{\rm top} ) \scal{ A + c} + \scal{A + c}^2
= {F + \Psi + \Phi}
\ee
Ref. \cite{BS} gives the interpretation of all fields in Eq.(\ref{basi}). 

To extend this 
horizontality condition and eventually define the vector symmetry, we found that we must make it compatible with reparametrization invariance, and, moreover, antighost dependent. 
 In fact, by finding the way of combining consistently 
 topological symmetry and reparametrization invariance, 
we will define scalar and vector topological invariances and reparametrization symmetry, with
 transformation laws that are globally well-defined. To obtain global consistency, we face
 the not so trivial question of expressing the transformation laws of
 Yang--Mills connections under reparametrizations, in the base space
 $M$, over which one compute the path integral. The appropriate
 language is well-known: it is the fiber bundle formalism. It allows one to define connections and their curvatures, and, eventually, 
 combine Eq.(\ref{basi}) with reparametrization symmetry. 
We will show that the symmetry transformations of the fields are
most easily obtained when they are lifted in the fiber bundle. Then
we will give the prescription to perform 
the projection on the base space, which defines the fields that one can insert in a path integral. Taking equal
to zero the background connection is basically the wish 
 for the impatient 
reader. The latter can identify the vector
ghost that expresses the reparametrization ghost in the base space with
its lifted expression in the fiber bundle. (The 
background connections for Lorentz and Yang--Mills invariances that we
will shortly introduce, $\bA$ and $\bomega$, define the horizontal
lift of the reparametrization ghost, from $\xi$ to $\xi^h$. Taking
$\bA=\bomega=0$ is often possible for field configurations that one
encounters in quantum field theory, so one can indeed often identify
$\xi$ and $\xi^h$) 
 
Using the fiber bundle language is not an unjustified excess of
mathematical rigor. It allows the construction of an action that is
a well-defined integral over the whole manifold, by ensuring that
the Lagrangian and the symmetry transformations involve globally
well defined objects. 

To define the reparametrization symmetry, one uses the notion of
spin--connection, with $\omega$ as a gauge field for the Lorentz
symmetry. This allows us to define the expression of combined
Lorentz and reparametrization symmetry, as was done long time ago in
the case of determining and classifying gravitational anomalies
\cite{mieg}\cite{marc}\cite{stora}. We will first consider the purely
gravitational bundle, and then generalize it for including the
topological Yang--Mills symmetry. 

Eq.(\ref{basi}) only involves the topological ghosts. In \cite{BS} the
antighosts are considered as a trivial BRST sector, which one
introduces in order to do the topological gauge-fixing. At the heart
of the notion of a TQFT, there is however a mapping between the ghost
and antighost Hilbert spaces. The introduction of the topological
vector symmetry will unexpectedly permit a transparent
geometrical interpretation of the antighosts, ``dual'' to that of the
ghosts, with some relationship to the idea of antiBRST symmetry.

\subsubsection{Pure gravitational case}
We will 
construct a gravitational ``horizontality condition" for defining the reparametrization symmetry and the way Yang--Mills connections and their topological ghosts transform under reparametrization. We will eventually 
reach an algebra that is globally well defined. 

To carry out this
program, we define the gravitational horizontality condition on
the total gravitational principal bundle $\Pi$ over the manifold
$M_n$, ($n$ is either 4 or 8), over which we will define the path
integral. \bea SO(n) \rightarrow & \Pi & \CR 
 & \downarrow& \CR
 & M& \eea
Then, we will introduce a relevant background Lorentz connection
$\bomega$.

A given connexion on $\Pi$ is equivalent to the selection of a decomposition
of its tangent space
\be T\Pi \cong TV\oplus TH \ee
It is known that the $\mathfrak{so}(n)$ valued $p$-forms on $M$ are
identified by the use of local trivializations to the equivariant forms in
$\mathfrak{so}(n) \otimes \Lambda^p TH^\ast$. The gauge potential defined on open sets of $M$ is the local trivialization of the
globally well defined connection in $\mathfrak{so}(n) \otimes TV^\ast$. 
In order not add too much notations, we will use the same
notations for the objects defined on $\Pi$ as for their local trivializations 
on $M$. 

The following covariant horizontality condition on $\Pi$ defines
a nilpotent and consistent graded differential operator $
\mathcal{S}$, (which we donnot yet interpret), acting on the connection $\omega$ and its ghost $\Omega$
\be\label{lbjt}
(d+ \mathcal{S}) \scal{ \omega + \Omega} + \scal{ \omega + \Omega}^2 =
\exp ({i_{\xi^h}}) R \ee

$\xi^h$ is the horizontal lift on $TH_0$ of the reparametrization ghost
vector field $\xi$ defined on $TM$. Eq.~(\ref{lbjt}) is the
generalization of that first found in \cite{mieg} for $M$. It is
aimed to determine transformations that contain local Lorentz
transformations and reparametrization transformations, in the BRST
formalism. The ${\xi^h}$- dependence, instead of a genuine ${\xi
}$-dependence, involves the existence of a background 
connexion $\bomega$, which allows us to make reparametrization
explicitly compatible with (Lorentz) gauge transformations\footnote{We
 note $TH_0$ the horizontal tangent space defined by the background
 connexion $\bomega$.}. $R= d_\omega
\omega$ is the definition of the curvature in $\Pi$.

The contraction operator
${i_{\xi^h}} $ acts on all relevant objects in $\Pi$, forms and
connections. An easy computation gives the following identity on
$\Pi$ : 
\be\label{typ}
\exp ( {-i_{\xi^h}}) ( d + \mathcal{S}) \exp( {i_{\xi^h}}) = d + \mathcal{S} -
\L_{\xi^h} + i_{ ({\mathcal{S} \xi^h} -{ \frac{1}{2} \{\xi^h, \xi^h \}} ) }
\ee
where $\L_{\xi^h}= [ {i_{\xi^h}} ,d] $.

The nilpotency of the graded operator
$( d + \mathcal{S})$ amounts to that in the rhs of
Eq.(\ref{typ}). This equation implies the introduction in $M$ of a
vector field $\varphi = 
\mathcal{S} \xi - \frac{1}{2} \{\xi, \xi\}$, that we may call a
ghost of ghost of the reparametrization ghost
$\xi$. 
We must have for consistency the following transformation laws : 
\bea\label{weyl}
\mathcal{S} \xi &=& \varphi + \frac{1}{2} \{\xi, \xi\} \CR
\mathcal{S} \varphi &=& \L_\xi \varphi
\eea 
(Ref. \cite{marc} explains the algebraic details of this construction). Provided that the later equation is satisfied, nothing forbids that $\varphi\neq 0$.

$ \varphi $ is a vector field on $TM$, with horizontal lift $\varphi
^h$ on $TH_0$ whose possible existence is, for the moment, just a
logical possibility.

 In ordinary gravity, in order to interpret
$\mathcal{S}$ as the BRST operator of plain reparametrization invariance, $\varphi$ must be fixed to zero. In this case, $\mathcal{S}$ just express the ordinary gravitational and Lorentz BRST symmetry. Formally, when $ \varphi \neq 0 $, Eq.(\ref{weyl}) looks like the Weyl extension of the Lie algebra of diffeomorphisms.

Since $\bomega$ is a background field, $\mathcal{S} \bomega = 0$ and the
BRST operator must commute with the horizontal lift that it defines,
$\mathcal{S} (\xi^h) = (\mathcal{S} \xi)^h$, that is
\bea
\label{geo}
\mathcal{S} \xi^h &=& \varphi^h + \frac{1}{2} \{\xi, \xi\}^h \CR
&=& \varphi^h + \frac{1}{2} \{\xi^h, \xi^h\} + \Scal{\frac{1}{2}
 i_{\xi^h}^{\,2} \bR}^v
\eea
where ${\left(\frac{1}{2} i_{\xi^h}^{\,2} \bR\right)}^v \equiv r^v$ is
the fundamental 
vertical vector associated to the $ \mathfrak{so}(n)$-valued element
$\frac{1}{2} i_{\xi^h}^{\,2} \bR$. Eq.(\ref{geo}) can be read as a
definition of the background curvature $\bR \equiv d_{\bomega} \bomega$. 

We can thus rewrite Eq.(\ref{typ}) in $\Pi$ under the following form :
\be
\exp({-i_{\xi^h}}) ( d + \mathcal{S}) \exp ({i_{\xi^h}}) = d + \mathcal{S} -
\L_{\xi^h} + i_{r^v} + i_{\varphi^h} 
\ee
We redefine on $\Pi$, $\t\Omega \equiv \Omega - i_{\xi^h}
\omega$, which must be written on $M$ as follows, by the use of local
trivialization
\be 
\t\Omega = \Omega - i_\xi (\omega - \bomega)
\ee
Indeed $\omega - \bomega$ is a tensorial form, and truly corresponds to a horizontal form on
$\Pi$; we can thus finally rewrite the gravitational horizontality condition in $\Pi$ as :
\be
(d + \mathcal{S} - \L_{\xi^h} + i_{r^v} + i_{\varphi^h} ) \scal{\omega
 + \t\Omega} + \scal{\omega+ \t\Omega}^2 = R \label{hpb}
\ee

 Eq.(\ref{hpb}) can be expanded in ghost number, and projected on $M$.
 
 For $\varphi=0$, the resulting pure
 gravitational
 transformation laws 
 depends on $\bomega$, and are as those that were computed in \cite{stora}, by
 asking that the gravitational BRST equations
 correspond to a Lie algebra.

 We will actually generalize Eq.(\ref{hpb}), when the relevant new ingredients will be introduced to combine it with the Yang--Mills topological symmetry and obtain the topological vector symmetry, with $\varphi\neq 0$. We will perform the projection at this moment.
So, we momentarily leave Eq.(\ref{hpb}) as it, and spend a few lines to comment on the operators that appear in it.

Since it is defined on $\Lambda^{\bullet} TV^\ast$, the contraction
operator ($i_{r^v}$) 
acts non trivially only on the connection. It 
 generates a term $\frac{1}{2} i_{\xi^h}^{\,2} \bR $ when one
expands in ghost number the horizontality 
condition.

$\L_{\xi^h}$ is defined as, 
$\L_{\xi^h} \equiv [i_{\xi^h}, d]$, where $d$ is the
exterior derivative on $\Lambda^{\bullet} T\Pi^\ast$. It is 
 defined for any $p$-form $w$ to be :
\be
\L_{\xi^h} w \equiv \Scal{\frac{d\,}{dt} \phi^\ast_{{\xi^h}, \, t}
 w}_{|t=0}
\ee
where $\phi^\ast_{{\xi^h}, \, t}$ is the pullback application of the flow
$\phi_{{\xi^h}, \, t}$, defined by 
\bea
\frac{d\,}{dt} \,\phi_{{\xi^h}, \, t}(p) &\equiv&
{\xi^h}_{|\phi_{{\xi^h},\,t}(p)} \CR 
\phi_{{\xi^h}, \, 0}(p) &\equiv& p
\eea
The curve $t \in [0, 1] \rightarrow \phi_{{\xi^h}, \, t}(p) \in \Pi$ is
 the horizontal lift of $t \in [0, 1] \rightarrow \phi_{\xi,
 \, t}(\pi p) \in M$ starting from $p$ , where $\pi$ is the
 projection from $\Pi$ to $M$ of the fiber bundle. 
As such, $\L_{\xi^h}$ is a parallel transport generator and since the
parallel transports preserves the tensoriality property of 
forms, $\L_{\xi^h}$ does also. It follows that the projection of 
$\L_{\xi^h}$ in $M$ must be locally expressed as: 
\be
\L_\xi + \delta_{\mathrm{Lorentz}}(i_\xi \bomega)\ee
In fact, the projection in $M$ of $i_{\xi^h}$ is $i_\xi$ for
a tensorial form, 
and $i_\xi (\omega - \bomega)$ for a
connection.

 We will now address the possibility $\varphi\neq 0$ by coupling gravity to Yang--Mills topological symmetry, so that 
 $\mathcal{S}$ will have a more general interpretation, which will
 allow us to define the vector topological symmetry.

\subsubsection{ Yang--Mills coupled to gravity}
In order to couple the Yang--Mills symmetry with gravity and obtain a horizontality condition for the 
topological Yang--Mills symmetry coupled to reparametrization
symmetry, we introduce another (Yang--Mills) principal
bundle $P$ :
\bea G \rightarrow & P & \CR
 & \downarrow& \CR
 & M& \eea
The additional horizontal Yang--Mills lift is defined by introducing a background
connexion $\bA$ on $P$.
One defines in $P$ : 
\be
( d + \mathcal{S} ) \scal{ A + \mathcal{C}} + \scal{A + \mathcal{C}}^2
= \exp ({i_{\xi^h}}) \scal{F + \Uppsi + \Upphi}
\ee
In an analogous way as in the previous section, we do a
left-multiplication by the operator 
$\exp ({-i_{\xi^h}})$, and we obtain: 
\be
( d + \mathcal{S} - \L_{\xi^h} + i_{f^v} + i_{\varphi^h} ) \scal{ A +
 \t{\mathcal{C}}} + \scal{A + \t{\mathcal{C}}}^2 
= F + \Uppsi + \Upphi
\ee
where $f^v \equiv \Scal{\frac{1}{2} i^{\,
 2}_{\xi^h} \bF}^v $ ( $\bF \equiv d_{\bA} \bA$) and $\t{\mathcal{C}}
\equiv \mathcal{C} - i_{\xi^h} A$.

To absorb the reparametrization ghost dependence, it is convenient to define a new operator $\hat{\mathcal{S}}$, from ${\mathcal{S}}$ \cite{marc}. For 
 all fields, but the Faddeev-Popov ghosts $\t\Omega$ and
$\t{\mathcal{C}}$, we define 
\be\label{gui}
\hat{\mathcal{S}} \equiv \mathcal{S} - \L_\xi -
\delta_{\mathrm{Lorentz}}(i_\xi \bomega) -
\delta_{\mathrm{gauge}}(i_\xi \bA); 
\ee
and for $\t\Omega$ and
$\t{\mathcal{C}}$, we define :
\bea\label{guib}
\hat{\mathcal{S}} \t\Omega \equiv \mathcal{S} \t\Omega - \bLc_\xi
\t\Omega + \frac{1}{2} i^{\, 2}_\xi \bR \CR
\hat{\mathcal{S}} \t{\mathcal{C}} \equiv \mathcal{S} \t{\mathcal{C}} - \bLc_\xi
\t{\mathcal{C}} + \frac{1}{2} i^{\, 2}_\xi \bF, 
\eea
where $\bLc_\xi \equiv [ i_\xi, d_{\bomega + \bA}]$. 

As a consequence of ${\mathcal{S}}^2=0$, one can 
 check that :
\be
\hat{\mathcal{S}}^2 = 
 - \L_\varphi -
\delta_{\mathrm{Lorentz}}(i_\varphi \bomega) -
\delta_{\mathrm{gauge}}(i_\varphi \bA)
\ee
and 
\be
\hat{\mathcal{S}} d+d\hat{\mathcal{S}}=0.
\ee

Using $\hat{\mathcal{S}}$ or ${\mathcal{S}}$ is a matter of convenience, which depends on the problem at hand.

 The ``decoupled" (i.e, with no explicit $\xi$-dependence) horizontality conditions read on
 $\Pi$ and $P$ :
\bea\label{magic}
(d + \hat{\mathcal{S}} + i_{\varphi^h} ) \scal{\omega
 + \t\Omega} + \scal{\omega+ \t\Omega}^2 &=& R \CR
( d + \hat{\mathcal{S}} + i_{\varphi^h} ) \scal{ A +
 \t{\mathcal{C}}} + \scal{A + \t{\mathcal{C}}}^2 
&=& F + \Uppsi + \Upphi \label{chapute}
\eea

They look almost as standard equations in flat space, except for the appearance of the operator $ i_{\varphi^h}$.

To summarize, we started from horizontality equations that are well-defined in the fiber bundle. By projection on the manifold $M$, we obtain 
 graded equations that determine the operator ${\mathcal{S}}$ in
 local coordinates, with transformation laws that are by construction globally well-defined, and will be shortly displayed. The redefinition of 
 $ \mathcal{S}$ into $ \hat{\mathcal{S}}$ gives simple expressions.
 
By expansion in ghost number, the later equations (\ref{magic}) determines the action of the 
BRST operator $\hat{\mathcal{S}}$ that is equivariant with respect to
the reparametrization group. After projection on $M$, they are:
\bea
 {\mathcal{S}} A - i_\xi F + d_A i_\xi \scal{ A - \bA} + d_A \t{\mathcal{C}} &=& \Uppsi\\
 {\mathcal{S}} \t{\mathcal{C}} -\bLc_\xi \t{\mathcal{C}} + \frac {1}{2}
 [\t{\mathcal{C}},\t{\mathcal{C}} ] + i_\varphi \scal{A-\bA} +
\frac {1}{2} i^{\, 2}_\xi \bF &=& \Upphi
\eea
We have similar equations for the action of ${\mathcal{S}}$ on 
$\omega$ and $\hat \Omega$. Using the relation between ${\mathcal{S}}$ and $\hat {\mathcal{S}}$, the equations can be equivalently rewritten in the following way, which shows more explicitly, term by term, that we have truly reached a globally well-defined definition of the BRST operation $\hat{\mathcal{S}}$:
\bea
\hat{\mathcal{S}} A +d_A\t{\mathcal{C}}
&=&\Uppsi\\
\hat{\mathcal{S}}\t{\mathcal{C}} + \frac {1}{2}
[\t{\mathcal{C}},\t{\mathcal{C}} ] + i_\varphi \scal{A -\bA } &=&\Upphi
\eea

This later expression of the symmetry is particularly convenient, in particular for computing the invariant Lagrangian.

\subsubsection{ Putting equal to zero the background connections }

The formula are simplest when one sets to zero the background connections $\bA=0$ and $\bomega=0$, and make no distinction between the vector fields in the fiber bundle and in $M$
(which is generally an improper choice from a global point of view,
but is sufficient in perturbative quantum field theory
around the trivial vacuum). This gives the transformation laws as
in \cite{marc}, which 
express the reparametrization and Yang--Mills symmetry. As indicated
at the time, they express the symmetry in two 
equivalent ways, which are sufficient to control the symmetry of the TQFT :
\bea\label{usefull}
 {\mathcal{S}} A +d_A \mathcal{C}
&=& \Uppsi + i_\xi F\CR
 {\mathcal{S}} \mathcal{C} +\frac {1}{2} [\mathcal{C},\mathcal{C} ]
 &=& \Upphi + i_\xi \Psi + \frac {1}{2} i^{\, 2}_\xi F 
\eea
and 
 \bea\label{useful}
( {\mathcal{S}} - \L_\xi ) A 
 +d_A\t{\mathcal{C}}
=\Uppsi
\CR
( {\mathcal{S}}-\L_\xi ) \t{\mathcal{C}}
+ \frac {1}{2} [\t{\mathcal{C}},\t{\mathcal{C}} ] + i_\varphi A 
 =\Upphi
\eea
where $\t{\mathcal{C}}=\mathcal{C} -i_\xi A$. The expression of ${\mathcal{S}} $ in Eq.(\ref{usefull}) is explicitly covariant, but tedious to use. The expression of $\hat{\mathcal{S}} \equiv ( {\mathcal{S}}-\L_\xi ) $ in Eq.(\ref{useful}) is convenient, 
 all dependence in $\xi$ is hidden, owing to the field redefinition
 $\mathcal{C} \to \mathcal{C} -i_\xi A$. Moreover, for an integral
 over the manifold, 
 $\mathcal{S}$ and $\hat{\mathcal{S}}$ invariances are the same. This
 field redefinition looks not globally well-defined, as $i_\xi A$ is
 ambiguous from a global point of view, but the above analysis has
 taught us how to remedy this, it must be understood as $ \mathcal{C}
 \to \mathcal{C} -
 i_\xi (A-\bA)$, and for the rest one should keep in mind the
 dependence in the background connection $\bA$, indicated in
 Eqs.(\ref{gui},\ref{guib}).

We actually have a consistent recipe: global
consistency is obtained from the simplest formulation with no background connection, provided
one replaces all connections that may appear in the form $i_\varphi
\omega$ or $i_\varphi A$ by their
difference with a background connection $i_\varphi \scal{A - \bA}$ or
$i_\varphi \scal{\omega - \bomega}$. This eventually defines a differential 
 $\hat{\mathcal{S}} $, which encodes the relevant information on the gauge symmetry and reparametrization. $\hat{\mathcal{S}} $-invariance defines the theory.

The above presentation makes a bridge between the facts that the 
expression for the BRST symmetry is simplest in the fibers bundle $\Pi,P$, defined over the manifold $M$, while it needs more elaborate formula on $M$, where quantum field theory is computed. It yields unambiguously the dependence in the background connection that delivers well-defined integrals over $M$. 

\subsection{Extended horizontality condition for the scalar and vector topological symmetry }
\label{HC}

We now
reach the important point of the paper, that, given a given covariantly constant vector $\kappa$ on $M$, we can geometrically build the topological vector BRST operator
out of a globally well-defined operator $ \hat{\mathcal{S}}$. the
vector symmetry will be shortly defined as a differential graded
operator $\delta$ with ghost number -1. We understood in section 2
that a BRST-exact action can possibly define a vector symmetry that
leaves it invariant. 
The question is to find a geometrical way of building this vector symmetry.

Since we have in mind the determination of the vector symmetry from an
``extended horizontality condition", we may wish to get a hint that
it is possibly contained in the geometrical formalism. We can see it
from the following indirect argument, which heuristically provides
evidence that the ordinary horizontality condition of a $\mathcal{N}_T
=~1$ TQFT contains the germs of another symmetry than the usual
topological BRST symmetry. There are in fact topological theories
with more than one scalar operator that can be 
identified to a BRST operator. They are known as balanced topological
theories \cite{moore}. They are often obtained by dimensional
reduction of a $\mathcal{N}_T =~1$ TQFT. Such theories have a
symmetry between 
ghosts and anti-ghost which is $SL(2, \mathds{R})$ in the case of two
charges, $\mathcal{N}_T = 2$. Both BRST algebras can be described
by a BRST-antiBRST horizontality condition, which displays a 
symmetry between ghosts and anti-ghosts.
 For instance, $\mathcal{N}_T = 2$ occurs when one
dimensionally reduces from 4 to 3 (or from 8 to 7) dimensions the
genuine d=8 (or d=4) $\mathcal{N}_T = 1$ topological Yang--Mills theory. In this case, the
topological ghost and antighost of the dimensionally reduced gauge field are symmetrical pair of
anticommuting vectors that belong to the fundamental representation
of $SL(2, \mathds{R})$, and the scalar topological ghost of ghost, the corresponding 
 antighost of antighost and the Higgs field, which results from the
dimensional reduction of the Yang--Mills field, fall into the adjoint
representation of $SL(2, \mathds{R})$. (Of course this phenomenon is related to the property of the R-symmetry in supersymmetric theories). 
In these cases, dimensional reduction allows us to obtain a BRST-antiBRST symmetry from a theory that seems to have only one BRST symmetry. 

Dimensional reduction occurs by giving a special role to a given dimension, and results in the elimination of the non--zero modes along 
a one-dimensional space $H$. 

 By enforcing reparametrization invariance in the relevant way, we will find, that on a reducible manifold $M \cong H \times N$
 where $H \cong \mathds{R}\ {\rm or} \ S_1 $, we can construct an extended horizontality condition 
 for 
 $\mathcal{N}_T = 1$ topological theories.
 An important point is that the invariant action will not depend on the constant vector $\kappa$ that defines the one-dimensional space $H$. This eventually defines the vector topological symmetry, which completes the ordinary scalar BRST symmetry, and shows that the 
 $\mathcal{N}_T = 1$ theories contain an enlarged symmetry. 

For the case of
$\Omega$ backgrounds, \label{reduced} we will need the existence of a Killing vector in addition to that of a covariantly constant vector field on the
 manifold. Eventually, we will obtain a twisted version of a deformed supersymmetry. 
 
 In what follows, we thus consider a manifold $M$ that contains at least a constant vector. This property reduces to the reducibility property for a simply connected manifold \cite{deRham}. (Reducibility only holds locally in the
 general case.)

\subsubsection{Obtaining of the extended horizontality condition }

We start from the formalism that introduces ghosts and antighosts of the Yang--Mills TQFT in a fully symmetrical way. We will shortly break this symmetry.

 The 
Lorentz invariant
 Yang--Mills BRST-antiBRST horizontality condition is
\be
( d + s + \bar s) \scal{ A + c + \bar c} + \scal{ A + c + \bar c }^2 =
F + \Psi + \bar \Psi + \Phi + \bar\Phi +L \label{gag}
\ee
It must be completed with its Bianchi identity that determines the action of 
$s $ and $ \bar s$ on the topological ghosts that occur on the right hand side, and ensures $(d + s + \bar s )^2=0$.
For such equations one has a conserved grading made of the sum of the ghost
number and the form degree on $M$. The total ghost number of 
$A, c, \bar c, \Psi, \bar \Psi, \Phi, L, \bar \Phi$ are respectively 
$0, 1, -1, 1, -1$, $2, 0, -2$, and their form degree $1, 0, 0, 1, 1$, $0, 0, 0$. A ghost antighost bigrading exists,
such that its values for $A, c, \bar c, \Psi, \bar \Psi, \Phi, L, \bar
\Phi$ are respectively 
$(0,0) (1,0), (0,1), (1,0), (0,1), (2,0),(1,1),(0,2)$. The scalar BRST and
antiBRST operators $s $ and $ \bar s$ have bigradings $ (1,0)$ and
$ (0,1)$, respectively. The net ghost number of a field with ghost
bigrading $(g,g')$ is simply $g-g'$. 

The fields $\bar\Psi$ and $ L$ do not show up in the $\mathcal{N}_T = 1$ theory, while the antiselfdual $2$-form antighost $\chi^{-1}_2$ and the scalar ghost $\eta^{-1}$ of this theory does not appear in Eq.(\ref{gag}). We will fill this apparent contradiction and 
come to the point of directly determining both $s$ and $\delta$
symmetries.

We classically couple the topological theory to
gravity, to express reparametrization invariance in $M$, but use the
freedom of introducing a vector $\varphi\neq 
0$, as generally shown in the last section. 
This will produces a symmetry operator $\hat{\cal S}$, which
obviously contains more information than the usual scalar operator
$s$. 

The existence of a covariantly constant vector field 
$\kappa$ on the manifold basically permits one to gauge-fix the
component $i_\kappa 
e^a$ to $\delta^a_0$. This property allows one to 
bypass the usual gravitational horizontality condition of the
vielbeins, which imposes $\varphi$ to be null in a fully $SO(n)$
invariant theory. Some deformations of the BRST transformation of $\xi$
can in fact be consistent with the closure of $\mathcal{S}$ on
$\omega$ and $\Omega$. The challenge is that the deformation must be
compatible with the Bianchi identity: 
\be
\mathcal{S}^2 \omega = i_\varphi R \hspace{10mm} \mathcal{S}^2 \Omega
= i_\xi i_\varphi R
\ee
Here, it is solved by some restriction of the geometry, such that
the equation of motion of the first order formalism $T = 0$, 
gives an $SO(n-1)$ holonomy curvature, leading to $i_\kappa R =
0$. This is because the gauge fixing of the vielbeins imposes to the
holonomy group of the second order curvature to be included in $SO(n-1)$.

 
We are thus allowed to give to
$\varphi$ a non-zero value in the $\kappa$ direction. 
To be definite, we choose 
$\varphi=-\kappa$.
The norm of 
$\kappa$ is an irrelevant quantity. Therefore, all 
identities must be homogeneous in $\kappa$. 
 It is equivalent to either
 impose the conservation of the bigrading $(g,g')$,
or to impose the conservation of the net ghost number
$g-g'$, assuming that $\kappa$'s bigrading is $(1,1)$. We conjecture that this
bigrading can be identified to the ghost antighost bigrading, in such a 
way that $g$ and $g'$ are both positive.

We identify $\hat{\cal S}= s^{\scriptscriptstyle (1,0)} +
s^{\scriptscriptstyle (0,1)}$, 
and $\t{\cal C}=c^{\scriptscriptstyle (1,0)} + c^{\scriptscriptstyle
 (0,1)} $.
The consistent horizontality
equation (\ref{chapute}) can be written as follows: 
\begin{gather}
( d + s^{\scriptscriptstyle (1,0)} + s^{\scriptscriptstyle (0,1)} -
i_{\kappa^h} )
\scal{ A + c^{\scriptscriptstyle (1,0)} + c^{\scriptscriptstyle
 (0,1)}} + \scal{ A + c^{\scriptscriptstyle (1,0)} + c^{\scriptscriptstyle
 (0,1)}}^2 \hspace{30mm}\CR
\hspace{40mm} = F + \Psi^{\scriptscriptstyle (1,0)} +
\Psi^{\scriptscriptstyle (0,1)} + \Phi^{\scriptscriptstyle (2,0)} +
\Phi^{\scriptscriptstyle (1,1)} + \Phi^{\scriptscriptstyle (0,2)}
\end{gather}

We now break the symmetry between the ghost and antighost sectors,
using the vector field $\kappa$. 
Each field of $(g,g')$ graduation must be homogeneous of degree $g'$ in
$\kappa$, and thus we define: 
\begin{gather}
c^{\scriptscriptstyle (0,1)} \equiv |\kappa| \bar c \hspace{10mm} 
\Psi^{\scriptscriptstyle (0,1)} \equiv g(\kappa) \eta + i_\kappa
\chi \CR 
\Phi^{\scriptscriptstyle (0,2)} \equiv |\kappa|^2 \bar\Phi
\end{gather}
where we defined the following $1$-form out of $\kappa$ :
\bea
g(\kappa) \equiv g_{\mu\nu} \kappa^\mu dx^\nu.
\eea
The $1$-form $g(\kappa) $ satisfies $i_\kappa g(\kappa) =
|\kappa|^2$, and will play an important role, together with the property
$(i_\kappa)^2=0$. A non zero value of $\Phi^{\scriptscriptstyle
 (1,1)}$ defines a consistent algebra; however there is no
corresponding invariant action. Therefore, we set $\Phi^{\scriptscriptstyle
 (1,1)} =0$.

 Of course, $c$ and $\bar c $ are identified as the Faddeev--Popov ghost and antighost, respectively.

The redefinition $\Psi^{\scriptscriptstyle (0,1)} \to (\eta, \chi)$
is $\kappa$ dependent in a non-trivial way. In fact, 
the decomposition of $\Psi^{\scriptscriptstyle (0,1)}$ implies that
the $2$-form representation of the holonomy 
group be reducible, in order that the pair $\eta^{-1} ,\chi^{-1}_{\mu\nu} $
has as many degrees of freedom as the vector
$\Psi^{\scriptscriptstyle (0,1)}_\mu$. In 8 dimensions, we thus 
suppose that $M$ has a holonomy group not larger than $Spin(7)$, so
that $\chi$ be 
antiselfdual in the octonionic point of view in eight dimensions, with
seven independent components (or 
 $\chi$ is antiselfdual in 4 dimensions, with three independent components).
Moreover, in order that $\kappa$ be
 globally well defined in 8 dimensions, the holonomy group must be
 included in $G_2$ (i.e, $N$ be a $G_2$-manifold in the reducible case).

We call:
\bea
s= s^{\gra{1}{0}}\CR
\delta=s^{\gra{0}{1}}
\eea

 Having introduced all these fields, we obtain\footnote{$g(\kappa)$
 must be seen on $P$ as $\pi^\ast 
 g(\kappa)$, where $\pi$ is the projection of the fiber bundle.} 
\begin{gather} 
\label{nero}
( d + s + \delta - i_{\kappa^h} ) \scal{ A + c + {|\kappa| \bar c} } + \scal{
 A + c + {|\kappa| \bar c}}^2 \hspace{40mm}\CR
\hspace{50mm} = F + \Psi + g(\kappa) \eta + i_{\kappa^h}
\chi + \Phi + |\kappa|^2 \bar\Phi
\end{gather}
and the associated Bianchi relation
\begin{gather}\label{bian}
( d + s + \delta - i_{\kappa^h} ) \scal{F + \Psi + g(\kappa)
 \eta + i_{\kappa^h} \chi + \Phi + |\kappa|^2 \bar\Phi}
\hspace{30mm}\CR \hspace{25mm} + \,[
A + c + {|\kappa| \bar c}\,,\, F + \Psi + g(\kappa) \eta + i_{\kappa^h} \chi +
\Phi + |\kappa|^2 \bar\Phi] = 0
\end{gather}

The property $( d + s + \delta - i_{\kappa^h} )^2=0$ implies in $M$:
\begin{gather}
\hat{\mathcal{S}}^2 = s^2 + \{s, \delta\} + \delta^2 = \L_\kappa +
\delta_{\mathrm{gauge}}(i_\kappa \bA) 
\end{gather}
We have therefore on all fields 
\begin{gather}\begin{array}
{rclcrcl}
s^2 &=& 0 \\*
 \{s, \delta\} &=& \L_\kappa +\delta_{\mathrm{gauge}}(i_\kappa \bA) \\*
\delta^2 &=&0 \\*
\end{array}
\end{gather}

And the decomposition in power of $\kappa$ of the transformation of
the reparametrization ghost implies
\begin{gather}\begin{array}{rclcrcl}
s \xi &=& -\frac{1}{2} \{\xi, \xi\} &\hspace{10mm}& \delta \xi &=&
- \kappa \\*
s\kappa &=& 0 & & \delta \kappa &=& 0
\end{array} 
\end{gather}

We will see shortly that a complete and finite field representation of the algebra can be found 
 in a consistent way in four and eight dimensions.

 \subsubsection{Resolution of the extended horizontality condition } 
The decomposition of the first equation (\ref{nero}) according to the
gradings gives: 
\begin{gather}\label{rclcrcl}\begin{array}{rclcrcl}
s A + d_A c &=& \Psi &\hspace{10mm}& \delta A + d_A |\kappa| \bar c &=&
g(\kappa) \eta + i_\kappa \chi \\*
s c + c^2 &=& \Phi & & \delta \bar c + |\kappa| \bar c^2 &=&
|\kappa| \bar\Phi 
\end{array}\CR
s |\kappa| \bar c + \delta c - i_\kappa \scal{A - \bA} + [c, |\kappa|
\bar c] = 0
\label{deltaA}\end{gather}
It is most convenient to use the $s$ and $\delta$ operators in the Cartan
representation defined by
\be
\label{delta2}
s_c \equiv s + \delta_{\mathrm{gauge}}(c) \hspace{10mm} \deltac \equiv
\delta + \delta_{\mathrm{gauge}}(|\kappa| \bar c)
\ee
On all fields, but $c$ and $\bar c$, one has:
\begin{gather}
s_c^2 = \delta_{\mathrm{gauge}}(\Phi) \hspace{10mm} \deltac^2 =
\delta_{\mathrm{gauge}}(|\kappa|^2 \bar\Phi) \CR
\{s_c, \deltac\} = \L_\kappa + \delta_{\mathrm{gauge}}\scal{i_\kappa
 A} 
\end{gather}
So, the decomposition of the Bianchi identity (\ref{bian}) gives 
\begin{gather}
s_c \Psi + d_A \Phi = 0 \hspace{10mm} \deltac \scal{g(\kappa) \eta +
 i_\kappa \chi} + |\kappa|^2 d_A \bar\Phi = 0 \CR
s_c \scal{g(\kappa) \eta + i_\kappa \chi} + \deltac \Psi - i_\kappa F
= 0 \CR 
\begin{array}{rclcrcl}
s_c \Phi &=& 0 &\hspace{20mm} & \deltac |\kappa|^2 \bar\Phi &=& 0 \\*
\deltac \Phi - i_\kappa \Psi &=& 0 & & s_c |\kappa|^2
\bar\Phi + |\kappa|^2 \eta &=& 0
\end{array}\end{gather}

Because of the bigraduation, the horizontality condition does not
fully determine the action of the $s$ and $\delta $ on all fields. 
Indeed, one has degenerate equations of the type 
$s$(antighost)$+\delta$(ghost)$=\ldots$. To raise the indetermination, we
introduce ``auxiliary'' fields, a scalar $b^0$ and a $2$-form $T_2^0$,
with : 
\be\begin{array}{rclcrcl}
s_c \chi &=& T & & s_c T &=& [\Phi, \chi] \\*
s_c \bar c &=& b & & s_c b &=& [\Phi, \bar c]
\end{array}\ee

It is by construction that we can consistently define the action of
$s$ and $\delta$ on the ``auxiliary'' fields. In practice, this
require a step by step computation, which yields the action of
$\delta$, such that 
 $\delta^2=0$, $\{ s, \delta\}= \Lc_\kappa $.

The field $\chi$ occurs in the horizontality relation
only trough its contraction along $\kappa$, $i_\kappa \chi$. Since
${i_\kappa}^2 = 0$, the resolution of equation is not yet established,
since $\chi$ is defined modulo terms that are $i_\kappa$-exact. 

A little of algebra must now be done to obtain the transformation of 
$\chi_{\mu\nu}$.
If we use the decomposition $\deltac =
\kappa^\mu s_\mu$, (which is only true locally), we have from
(\ref{deltaA}) and (\ref{delta2}) that
\be s_{[\mu} A_{\nu]} = - \chi_{\mu\nu} \hspace{10mm} s_{\{\mu}
s_{\nu\}} A_\sigma = - g_{\mu\nu} D_\sigma \bar \Phi \ee
This gives 
\be
s_\sigma \chi_{\mu\nu} = - s_\sigma s_{[\mu} A_{\nu]} \ee
and we deduce from the decomposition 
\begin{figure}[h]
\begin{center}
\begin{picture}(115,30)(-28,-5)
\put(-20,0){\framebox(10,10)}
\put(-5,0){\makebox(10,10){$\otimes$}}\put(10,5){\framebox(10.5,10)}
\put(10,-5){\framebox(10.5,9.5)} 
\put(30,0){\makebox(10,10){$=$}}
\put(50,10){\framebox(10,10)} \put(50,0){\framebox(10,9.5)}
\put(50,-10){\framebox(10,10)} 
\put(65,0){\makebox(10,10){$\oplus$}}
\put(80,0){\makebox(10,10){$2 \times$}}
\put(95,5){\framebox(10,10)} \put(95,-5){\framebox(10,9.5)}
\put(105.5,5){\framebox(10,10)} 
\end{picture}
\end{center}
\end{figure}

\hspace{-6mm}that 
\be
s_\sigma \chi_{\mu\nu} = 2 g_{\sigma[\mu} D_{\nu]} \bar\Phi -
s_{[\sigma} s_\mu A_{\nu]}
\ee
The ghost number and dimension of fields are such that, without 
introducing other fields, $s \chi$ must be proportional to $d_A
\bar \Phi$
\be
s_\sigma \chi_{\mu\nu} = 2 g_{\sigma[\mu} D_{\nu]} \bar\Phi -
{C^{\star}_{[\sigma\mu\nu]}}^\rho D_\rho \bar\Phi 
\ee

We have introduced an invariant tensor $C^{\star}_4$ on the manifold. (It
is the $\epsilon$ tensor for 
4-manifolds, and the octonionic self-dual 4-form for Joyce manifolds.)

We already know that the condition $\chi_{\mu\nu} =
{P^-_{\mu\nu}}^{\sigma\rho} 
\chi_{\sigma\rho}$, for this field to count for $n-1$ degrees of
freedom in $n$ dimensions. Since $ \deltac^2 \chi = [\bar\Phi, \chi]
$, $C^{\star}_{\mu\nu\sigma\rho}$ must be totally antisymmetric and
that 
\be
\Scal{ 2 \delta_{\mu\nu}^{\{\sigma|\kappa} -
 {C^{\star}_{\mu\nu}}^{\{\sigma|\kappa}}
{{P^{-\,\rho\}}}_\kappa}^{\theta\tau} = g^{\sigma\rho}
{P^-_{\mu\nu}}^{\theta\tau} 
\ee
which gives
\be 
{P^-_{\mu\nu}}^{\sigma\rho} = \frac{2}{n} \Scal{
 \delta_{\mu\nu}^{\sigma\rho} - \frac{1}{2}
 {C^{\star}_{\mu\nu}}^{\sigma\rho}}
\ee 
We know from \cite{BL} that the only possibilities to construct such
projector with the holonomy group reduced to at most the maximal
invariant subgroup of $SO(n)$ are in four and eight dimensions. This
is a further check of the need of a holonomy group of $M$. We will
adopt a unified notation 
for this projector in four and eight
dimensions, with the convention that $C$ is just the unity
 in four dimensions and the octonionic $4$-form in eight dimensions :
\be
w_2^- \equiv \frac{2}{n} \scal{ w_2 - \star C w_2}
\ee


 
\subsection{Expression of the $s$ and $\delta$ symmetries }

The resolution of the horizontality condition has thus given us the following 
expression for the action $s$ and $\delta$ on the fields we started from, plus the fields that we had to introduce to solve the degeneracies. These transformation laws are written in a way that is globally well defined:

\bea
\label{equipart}
&\begin{array}{rclcrcl}\label{impf}
s_c A &=& \Psi &\hspace{10mm}& \deltac A &=& g(\kappa)
\eta + i_\kappa \chi \\* 
s_c \Psi &=& -d_A \Phi & & \deltac \Psi
&=& i_\kappa \scal{ T+ F} + g(\kappa) [\Phi, \bar\Phi] \\* 
s_c \Phi &=& 0 & & \deltac \Phi &=& i_\kappa \Psi \\* 
&&&&&&\\*
s_c \bar\Phi &=& \eta & & \deltac \bar\Phi &=& 0 \\* 
s_c \eta &=& [\Phi, \bar\Phi] & & \deltac \eta &=&
\Lc_\kappa \bar\Phi \\* 
&&&&&&\\*
s_c \chi &=& T & & \deltac \chi
&=& \frac{n}{2} \scal{g(\kappa) d_A \bar\Phi}^- \\* 
s_c T &=& [\Phi, \chi] & & \deltac T &=& -\frac{n}{2}
\scal{g(\kappa) d_A \eta + g(\kappa) [\bar\Phi, \Psi]}^- + \Lc_\kappa
\chi \\*
&&&&&&\\*
\end{array}&
\\ 
&\begin{array}{rclcrcl}
s \,c &=& \Phi - c^2 & \hspace{10mm} & \delta\, c &=& i_\kappa
\scal{A-\bA} - |\kappa| b \hspace{50mm}\\* 
\CR
\label{trivpart}
s \bar c &=& b - [c, b] &\hspace{10mm} & \delta \bar c &=& |\kappa|
\scal{ \bar\Phi - {\bar c}^2} \\*
s b &=& [\Phi, \bar c] - [c, b]& & \delta b &=& \Lc_\kappa \bar c -
|\kappa| \eta \\*
\end{array}& \\
\eea

Let us give for the sake of clarity the explicit $c$ and $\bar c$ dependence in the transformation laws Eqs.~(\ref{impf})
\bea
\label{equipart}
&\begin{array}{rclcrcl}
s A &=& \Psi -d_A c &\hspace{10mm}& \delta A &=& g(\kappa)
\eta + i_\kappa \chi - |\kappa| d_A \bar c\\* 
s \Psi &=& -d_A \Phi -[c,\Psi ] & & \delta \Psi
&=& i_\kappa \scal{ T+ F} + g(\kappa) [\Phi, \bar\Phi] - |\kappa|[\bar c,\Psi ]\\* 
s \Phi &=& -[c,\Phi ] & & \delta \Phi &=& i_\kappa \Psi 
 -|\kappa| [\bar c, \bar \Phi ]\\* 
&&&&&&\\*
s \bar\Phi &=& \eta -[c,\bar \Psi ]& & \delta \bar\Phi &=& -|\kappa| [\bar c,\bar \Psi ] \\* 
s \eta &=& [\Phi, \bar\Phi] -[c, \eta ]& & \delta \eta &=&
\Lc_\kappa \bar\Phi -|\kappa| [\bar c, \eta ]\\* 
&&&&&&\\*
s \chi &=& T -[c, \chi ]& & \delta \chi 
&=& \frac{n}{2} \scal{g(\kappa) d_A \bar\Phi}^- -|\kappa| [\bar c, \chi ] \\* 
s T &=& [\Phi, \chi]-[c, T ] & & \delta T &=& -\frac{n}{2}
\scal{g(\kappa) d_A \eta + g(\kappa) [\bar\Phi, \Psi]}^- + \Lc_\kappa
\chi -|\kappa|[\bar c, T ] \\*
&&&&&&\\*
\end{array}&
 \eea

 The following remarks are important :


- As for an explicit dependence on the background connection, it only
occurs in $\delta c$.

 - The part of the BRST algebra that is relevant for untwisting
 only involve the fields in the first sector of the BRST equations, in 
 Eqs.~(\ref{impf}). It is important to note that the BRST invariance, introduces the Faddeev-Popov ghost in a way that is compatible with the $\delta$ invariance. This property is, in particular, very important for the questions relative to renormalization of supersymmetric gauge theories.

- In the flat space limit, we can define the anticommuting generators
$s_\mu$, using $\deltac = \kappa^\mu s_\mu$, and expanding in
$\kappa^\mu$, both in 4 and 8 dimensions. The action of $s_\mu$ for
the fields in (\ref{equipart}) 
 identifies itself with the vector symmetry \cite{Tet} used in
 $d=4$ and obtained by twisting the supersymmetry algebra. However,
 here, the presence of auxiliary fields and Faddeev--Popov ghosts and
 antighost fully ensures consistency.

- The fields $\bar c$ and $b$ are indispensable in
order to close the $s$ and $\delta$ operator on the Fadeev-Popov ghost
$c$. In fact, the existence of the $\delta$ symmetry, and its link with
reparametrization invariance, is likely to provide the geometrical
interpretation of 
$\bar c$ and $b$, as well as of other antighosts of the TQFT.

- The ghost
antighost duality which will be defined in the next section will make
the latter point more precise. 

- Power counting arguments based on the dimensionality and ghost
number of fields show that the above transformation laws are the most
general ones that solve the relations $s^2=\delta^2=0$, $\{s,\delta\}
=\bLc_\kappa$, up to irrelevant field multiplicative renormalization
factors and linear field redefinitions for the``auxiliary" fields $b$
and $T$. The geometrical horizontality equation that we defined in the
fiber bundle actually solve this constraint. 

In the next section, we will construct invariant actions for the
symmetry. Eventually, we will discuss this untwisting toward the
Poincar\'e superalgebra.

\section{The action}
\label{action}

\subsection{The topological/supersymmetric action as a $s$-exact term}

We now have a complete realization of the $s$ and $\delta$
symmetries, and wish to compute an invariant action. 

Even if the algebra is defined
only on a manifold of reducible tangent bundle, we demand a
$Spin(7)$ or an $SO(4)$ 
invariant action in respectively eight and four dimensions. This means
that the action must be independent of $\kappa$, which is an non trivial 
requirement. (Our construction
 needs a base space of the topological Yang--Mills theory that
 contains a constant vector. However, we wish the theory be generalizable
 for a more general 
 manifold, provided it has the required holonomy for defining
 self--duality equations.) 

We will focus on the terms of power counting corresponding to the
strictly renormalizable case in four dimensions, and extend this
requirement in 8 dimensions, which can be done by a formal power
counting argument (so we exclude higher order derivative terms).

The only way to construct an $\hat{\mathcal{S}}$-exact action, which
is invariant under gauge transformations, reparametrization invariant and independent
of $\kappa$, is such that
\bea\label{saction}
S = s \Uppsi^{\scriptscriptstyle (-1,0)} + \delta
\Uppsi^{\scriptscriptstyle (0,-1)} \CR
\delta \Uppsi^{\scriptscriptstyle (-1,0)} = 0 \hspace{10mm} s
\Uppsi^{\scriptscriptstyle (0,-1)}= 0
\eea
There is only one solution of $s\Uppsi^{\scriptscriptstyle (0,-1)}=0$
which gives an action independent of $\kappa$. In turn, the $\delta$-invariance completely
determines $\Uppsi^{\scriptscriptstyle (-1,0)}$ (with the hypothesis that there are nor higher order derivative terms). As a
matter of fact, up to a global scale factor and a
topological term, these two solutions give the same action. One has indeed :
\be s \Uppsi^{\scriptscriptstyle (-1,0)} = \delta
\Uppsi^{\scriptscriptstyle (0,-1)} + \frac{1}{2} \int_M C_{\, \wedge}
\trace F_{\, \wedge} F \ee 
So, we restrict ourself to the $\delta$ invariant gauge function, and we
define :
\be S = s \Uppsi \ee
where $\Uppsi$ is a $\delta$-closed gauge invariant functional, such that :
\be \label{uppsi} \Uppsi = \int_M \trace \Scal{ \chi \star \scal{ F +
 \frac{2}{n} T} + \bar\Phi d_A \star \Psi + \star \eta [\Phi,
 \bar\Phi]} \ee 
 
One has the usual interpretation that the action $ s \Uppsi $ is the gauge-fixing of a topological invariant. But the gauge fixing--function has been fixed from $\delta_{\bar c}$ invariance.

We will shortly show that there is a duality symmetry between $s$
and $\delta$, and that one can express this action as a
$s\delta$-exact action.

\subsection{Gauge-fixing part}

We identify $\bar c$ and $b$ as the familiar
Faddeev--Popov pair of an antighost and a Lagrange multiplier for
gauge-fixing the Yang--Mills invariance. We could use a term like $s(\bar c\partial\cdot A)$, which violates the $\delta$-invariance, but we prefer a $s
\delta$-exact term: 
\be
s \delta \frac{1}{2 |\kappa|} \int_M \trace \Scal{ \scal{ A- \bA}
 \star \scal{ A- \bA}} 
\ee

This gauge-fixing term breaks the $SO(n)$ invariance, since it depends on $\kappa$. From the
point of view of the equivariant theory, however, the relevant action
is the part of the action that is gauge invariant. It is determined by both $s$ and
$\delta$ invariance. A further $ SO(n)$ invariant gauge-fixing of
the yang--Mills 
symmetry implies the breaking of the vector symmetry. (This is of course a gauge-fixing artifact, which does not appear for the gauge invariant observables, which are $\kappa$-independent). This is
understandable in the untwisted formalism, where a supersymmetric
gauge-fixing process of the Yang--Mills invariance only holds in a
fully superspace version of the theory, and would yield an infinite
number of fields in 8 dimensions.\footnote{ A refinement of our work
 can be reached by extending our result in the context of equivariant
 invariance, which implies the introduction of a background gauge
 symmetry, along the line of \cite{5d,5d2}.}

\subsection{A $s \delta$-exact expression of the
 supersymmetric action} 
\label{sdexact} 
 
As in the case of $\mathcal{N}_T = 2$ topological theories, the action
(\ref{saction})
is $s \delta$-exact on a reducible manifold. Indeed, we can verify the following very suggestive formula, which shows directly $s$ and $\delta$ invariances :

\be
S = s \, \delta \int_M \usk \, \mathscr{F} \ee
with
\be 
\mathscr{F} = \trace \Biggl( \frac{1}{2} 
 g(\kappa)_{\,\wedge}C_{\,\wedge}\Scal{ \scal{ A- \bA}_{\,
 \wedge}\scal{ F + \bF} - \frac{1}{3} \scal{A- \bA}^3}
+ \scal{g(\kappa) \eta + {i_\kappa \chi}}_{\, \wedge}
 \star \Psi \Biggr) 
\ee 
where we recognize the last term to be the Chern--Simon term
\be
 d \, \trace \Scal{ \scal{ A- \bA}_{\, \wedge}\scal{ F + \bF } -
 \frac{1}{3} \scal{A- \bA}^3} = \trace \scal{ F_{\, \wedge} F -
 \bF_{\wedge } \bF} \ee
Note that in the case of a trivial fiber bundle we can take $\bA = 0$
and recover the ordinary form of the Chern--Simon term.
Unlike in the $\mathcal{N}_T = 2$ case, the ``potential''
$\int_M \mathscr{F}$ cannot be interpreted as a Morse function for the
theory. Indeed, 
the function $\int_M \trace \Scal{ \frac{1}{2}
 g(\kappa)_{\,\wedge} C_{\,\wedge}\scal{ A d A + \frac{2}{3} A^3
 }} $ 
 has degenerated
critical points, ($i_\kappa A$ does not appear in its equations of
motions), while the Chern--Simons potential gives a well defined
Morse function 
 in dimension $n-1$, when there is a balanced topological
theories \cite{moore}.

\subsection{Ghost antighost duality}

In spite of the fact that $\int_M \mathscr{F}$ is not a well defined Morse
function, we can interpret formally
\be F + g(\kappa) \eta + i_\kappa \chi + |\kappa|^2 \bar\Phi \ee
as a curvature as in the case $\mathcal{N}_T = 2$. 
This interpretation can be seen by introducing an 
involution $\ast$, which gives a duality between ghosts and
anti-ghosts.
\begin{gather}
\ast A = A \hspace{10mm} \ast T = - T -\frac{n}{\,\,2|\kappa|^2}
\scal{g(\kappa)i_\kappa F}^- \CR
\ast \Psi = g(\kappa) \eta + i_\kappa \chi \hspace{10mm} \ast \eta =
\usk \,i_\kappa \Psi \hspace{10mm} \ast
\chi = \frac{n}{\,\,2|\kappa|^2} \scal{g(\kappa) \Psi}^-\CR
\ast \Phi = |\kappa|^2 \bar\Phi \hspace{10mm} \ast \bar\Phi =
\usk \, \Phi \CR
\ast c = |\kappa| \bar c \hspace{10mm} \ast b = - b + [c, \bar c] +
\frac{1}{|\kappa|} i_\kappa \scal{ A - \bA} \hspace{10mm} \ast \bar c
= \frac{1}{|\kappa|} c
\end{gather}
This involution relates both operators $s$ and $\delta$, as follows :
\be
\ast s \ast = \delta \hspace{10mm} \ast \delta \ast = s
\ee
Note that $\ast$ does not act as a derivative but as a group element,
that is 
\be
\ast \scal{ A \cdot B} = \ast A \cdot \ast B
\ee

Indeed, by definition of $*$, we can associate to the $\delta$-invariant gauge function $\Uppsi$ defined in Eq.~(\ref{uppsi}) a mirror $s$-invariant gauge function, which defines a slightly different 
 $\delta$-exact gauge invariant action $*S$, which gives the same observables. By construction, $*S$ is 
$s$- and $\delta$-invariant. However $*S$ is $\kappa$-dependent,
but this dependence disappears after elimination of the auxiliary field
$T$, and then $*S$ becomes equal to $S$.

The horizontality relation gives indeed :
\begin{gather}
(d + \delta)\scal{ A+ |\kappa| \bar c} + \scal{ A+ |\kappa| \bar c}^2
= F + g(\kappa) \eta + i_\kappa \chi + |\kappa|^2 \bar\Phi \CR
(d_A + \deltac) \scal{ F + g(\kappa) \eta +
 i_\kappa \chi + |\kappa|^2 \bar\Phi} = 0
\end{gather}
The gauge function $\ast \Uppsi$ replaces the ordinary
gauge function $ \Uppsi$, when one interchanges the r\^oles of $s$ and
$\delta$: 
\be
\ast \Uppsi = \usk\,\int_M \trace
\biggl(- g(\kappa)_{\,\wedge}C_{\,\wedge} \Psi_{\,\wedge}F-
 g(\kappa) \Psi \star T + \Phi d_A \star \scal{g(\kappa)\eta +
 i_\kappa \chi} + \star i_\kappa \Psi [\bar\Phi, \Phi]\biggr)
\ee
Eventually, we can define the topological observables as functions of the dual variables, using any 
gauge invariant polynomial in the fields $P\scal{F,\Psi,\Phi}$, as follows :
\bea
\ast \left< P\scal{F,\Psi,\Phi} \right> &=& \ast \int\mu\,\,
P\scal{F,\Psi,\Phi} e^{-s \Uppsi} \CR
&=& \int\ast\mu \,\,P\scal{F,g(\kappa) \eta+ i_\kappa
 \chi,|\kappa|^2\bar\Phi} e^{-\delta \ast \Uppsi} \CR
&=& \left< P\scal{F,g(\kappa) \eta + i_\kappa \chi,|\kappa|^2\bar\Phi}
\right>
\eea
The descent equations are obtained by changing $d+s$ into $d+\delta$. As a conclusion, after the duality operation $*$, the curvature of the ``big bundle''
defined in \cite{Mo} becomes $ F + g(\kappa) \eta + i_\kappa
\chi + |\kappa|^2 \bar\Phi$ instead of $F + \Psi + \Phi$. 

\section{Conservation of the energy-momentum tensor
 antecedent~$\Lambda_{\mu\nu}$} 
We now verify that the operator $\delta$ is truly generated by the
conservation law of 
 the BRST antecedent of the energy momentum tensor. This
computation is tricky, since the topological action is
generally only invariant under $G \subset SO(n)$.
Let us write the action $s
\Uppsi$ as :
\begin{multline}
S = \int_M d^n x \sqrt{g} \, \trace
\biggl (\frac{1}{2}T^{\mu\nu}\scal{-F_{\mu\nu} + \frac{2}{n}
 T_{\mu\nu}} + \chi^{\mu\nu} \scal{D_\mu \Psi_\nu - \frac{1}{n}
 [\Phi,\chi_{\mu\nu}]} \biggr .\\*\biggl . + \eta D_\mu \Psi^\mu + \bar\Phi
 \{\Psi_\mu,\Psi^\mu\} - \bar\Phi D_\mu D^\mu \Phi + [\Phi,\bar\Phi]^2 -
 \eta[\Phi,\eta] \biggr )
\end{multline}
To compute the energy momentum tensor, we use the standard formula :
\be
T_{\mu\nu} = \frac{1}{e} \,\delta_{ab}\, 
 \frac{\delta S}{\delta e^\mu_b} \,e^a_\nu 
\ee 
We have to understand the way the projector on self-dual 2-forms
transform under variations of the vielbeins. 
In 8 dimensions, the variation of this projector $P^-$ with respect to the 
vielbeins is\footnote{In 4 dimensions, one replaces ${C_{ab}}^{cd}$ by 
the $\epsilon$ symbol}: 
\bea
\delta {P^-_{\mu\nu}}^{\sigma\rho} &=& -\frac{1}{n}\delta \scal{e_\mu^a e_\nu^b
 e^\sigma_c e^\rho_d} {C_{ab}}^{cd} \CR
&=& \frac{2}{n} {C_{\lambda[\nu}}^{\sigma\rho} e^a_{\mu]} \delta
e_a^\lambda - \frac{2}{n} {C_{\mu\nu}}^{\lambda[\rho} e^a_\lambda
\delta e_a^{\sigma]}
\eea
Using this formula, the energy momentum tensor is given by 
\bea
T_{\mu\nu} &=& \trace \Biggl ( - {T_\nu}^\sigma F^+_{\mu\sigma} + 2
{\chi_\nu}^\sigma D_{[\mu} \Psi_{\sigma]_+} -2 D_{\{\mu} \eta \cdot
\Psi_{\nu\}} + 2\bar\Phi\{\Psi_\mu,\Psi_\nu\} + 2D_{\{\mu} \bar\Phi
\cdot D_{\nu \}}\Phi 
\Biggr . \CR & & \hspace{1cm} + g_{\mu\nu}\biggl (
\scal{\frac{2}{n}-\frac{1}{2}} T^{\sigma\rho} \scal{-F_{\sigma\rho} +
 \frac{2}{n} T_{\sigma\rho}} + \scal{\frac{4}{n}-1} \chi^{\sigma\rho}
\scal{ D_\sigma \Psi_\rho - \frac{1}{n} [\Phi,\chi_{\sigma\rho}]}
\biggr . \CR & & \hspace{3cm} \Biggl . \biggl . + D_\sigma \eta \cdot
\Psi^\sigma - \bar\Phi \{\Psi_\sigma, \Psi^\sigma\} - D_\sigma \bar\Phi \cdot 
D^\sigma \Phi - [\Phi, \bar\Phi]^2 + \eta [\Phi, \eta] \biggr ) \Biggr ) \CR
\eea

This tensor is not
symmetric in eight dimensions and its antisymmetric part is
antiselfdual (it is valued in $\mathfrak{so}{\textstyle(8)}
\backslash \mathfrak{spin}{\textstyle(7)}$). This is allowed to the
fact that only isometries which preserve the octonionic $4$-form $C$
define conserved charges. In four dimensions, it is
symmetric. 

Since the BRST operator does not depend on $e^a_\mu$, we have :
\be
T_{\mu\nu} = s \, \frac{1}{e} \,\delta_{ab}\, 
 \frac{\delta \Uppsi}{\delta e^\mu_b} \,e^a_\nu 
\ee 
In this way, we find a $s$-antecedent of the energy momentum tensor :
\bea 
\Lambda^{\scriptscriptstyle (0)}_{\mu\nu} &=& g_{\mu\nu} \trace \Biggl
( \,\, \frac{1}{2} 
\scal{\frac{4}{n} - 1} 
\,\chi^{\sigma\rho}\Scal{-F^-_{\sigma\rho} + \frac{2}{n} T_{\sigma\rho}}
 +\Psi^\sigma D_\sigma \bar\Phi - \eta [\Phi, \bar\Phi]
\Biggr ) \CR 
& & \hspace{4cm} - \trace \Scal{{\chi_{\nu}}^\sigma F^+_{\mu\sigma} +2
 \Psi_{\{\mu} D_{\nu\}} \bar\Phi} 
\eea
It is not yet conserved. However, by adding a $s$-exact term to
$\Lambda^{\scriptscriptstyle (0)}_{\mu\nu}$, we can enforce the
conservation law, as follows : 
\bea
\nabla^\nu \Lambda_{\mu\nu} &\equiv& \nabla^\nu \Biggl(
\Lambda^{\scriptscriptstyle (0)}_{\mu\nu} - s\,\trace\Scal{ \frac{1}{2}
 {C^{\star}_{\mu\nu}}^{\sigma\rho} F_{\sigma\rho} \bar\Phi +
 \chi_{\mu\nu} \eta - \frac{1}{2} \scal{1 -
 \frac{4}{n}} {\chi_{[\mu}}^\sigma \chi_{\nu]\sigma}}\Biggr) \CR
&=& -\trace \Biggl ( \scal{-g_{\mu\nu}
 \eta + \chi_{\mu\nu}} \frac{\delta^L S}{\delta A_\nu} + \scal{F_{\mu\nu} - T_{\mu\nu}
 + g_{\mu\nu} [\Phi,\bar\Phi]} \frac{\delta^L S}{\delta \Psi_\nu} - \Psi_\mu
\frac{\delta^L S}{\delta \Phi} \Biggr .\CR 
& & \hspace{20mm} +
D_\mu \bar\Phi \frac{\delta^L S}{\delta \eta} - n D^\nu \bar\Phi \frac{\delta^L
 S}{\delta \chi^{\mu\nu}} + \frac{n}{2}
\scal{ 2\,D^\nu \eta + D_\sigma \chi^{\sigma\nu} - 2\, [\bar\Phi,\Psi^\nu]}
\frac{\delta^L S}{\delta T^{\mu\nu}} \CR
& &\Biggl .\hspace{35mm} - \frac{n}{2} \chi^{\mu\sigma} D^\nu
\frac{\delta^L S}{\delta T^{\nu\sigma}} + \scal{\, \frac{n}{2} - 2} D^\nu
\Scal{{\chi_\mu}^\sigma \frac{\delta^L S}{\delta T^{\nu\sigma}}}
\Biggr ) \CR
\eea
From the above equation we can recover 
the explicit form of the 
functional generators $V_{A\, \mu}$ that we introduced in section \ref{VS}. 
We can verify that the resulting symmetry
truly correspond to the $\delta$-operator 
that we build in section \ref{HC} directly from the horizontality condition.


\section{Untwisting toward Yang--Mills supersymmetry}
The theories that we have determined in four and eight dimensions from $\delta$ and $s$ invariances
correspond by twist to superYang--Mills $\mathcal{N} = 2$ theories. 
 The equivariant form of the scalar and the $\delta$ BRST operators
can in fact be identified on twisted combinations of spinorial supersymmetry generators. As a matter of
fact, if we define the theory on a manifold that is 
sufficiently constrained to admit two supersymmetries of opposed
chirality, the equivariant form of these $s$ and $\delta$ can be related to supersymmetry transformations as follows:
\be
\theta s_c = \delta_{\mathrm{susy}}(\theta \zeta) \hspace{10mm} \theta \deltac = \delta_{\mathrm{susy}}(i\theta \, \ba \kappa \zeta)
\ee
for $\theta$ an anticommuting parameter and $\zeta$ a chiral covariantly
constant spinor. As it is well known, in eight dimensions we can
construct the $\mathcal{N}=2$ superYang--Mills algebra from the dimensional
reduction on a torus of the ten dimensional superYang--Mills algebra,
with a further 
 Wick rotation of the $\mathcal{N} = 1$ theory that is generally
 defined on Minkowski space. The transition from 
$\mathcal{N} = 1$ to $\mathcal{N}= 2$ is allowed by the fact that the
Majorana-Weyl condition is consistent in eight dimensions for an
Euclidean space and not for a Minkowski one. The {\it Euclidean}
action obtained in 
this way is
\begin{multline}
\int_M d^8x \sqrt{g}\, \trace\biggl(-\frac{1}{4}F_{\mu\nu}F^{\mu\nu} +
\frac{1}{2} D_\mu \phi_1 D^\mu \phi_1 -\frac{1}{2}D_\mu\phi_2D^\mu\phi_2
 -\frac{i}{2} \left(\overline{\lambda}
 \baa D\lambda\right) \biggr.\\*\biggl.
+\frac{1}{2}[\phi_1,\phi_2]^2
 +\frac{1}{2}\left(\overline{\lambda}\gamma_9[\phi_1,\lambda]\right)
 -\frac{1}{2}\left(\overline{\lambda}[\phi_2,\lambda]\right) \biggr) 
\end{multline}
It is invariant under the following supersymmetries for any
covariantly constant spinor $\epsilon$\bea &\delta_{\rm {susy} }A_\mu &= -i(\overline{\epsilon} \gamma_\mu \lambda)\CR
 &\delta_{\rm {susy} } \phi_1 &= -(\overline{\epsilon} \gamma_9 \lambda)\CR
 &\delta_{\rm {susy} } \phi_2 &= -(\overline{\epsilon} \lambda)\CR
 &\delta_{\rm {susy} } \lambda &= \ba F\epsilon-i\gamma_9\baa D\phi_1\epsilon
-i\baa D\phi_2\epsilon +\gamma_9 [\phi_1,\phi_2]\epsilon \CR
\eea
If $M$ is defined to be a $Spin(7)$-manifold, it contains a chiral
covariantly constant spinor $\zeta$. We choose it Majorana-Weyl with
norm equal to one. One can further construct the 
octonionic $4$-form, as follows: 
\bea
 \scal{\overline{\zeta}\zeta} = 1 \hspace{2cm}
4!\scal{\overline{\zeta}\gamma_{\mu\nu\sigma\rho}\zeta} =
C_{\mu\nu\sigma\rho}
\eea
We can decompose the Majorana spinor
fields of the theory in term of differential forms, by projection over
$\zeta$, which is the definition of the twist in 8 dimensions:
\bea
\lambda_+ &=& \scal{\eta + \baa \chi} \zeta\CR
\lambda_- &=& i\, \baa \Psi \zeta 
\eea
Here $\eta$, $\chi$ and $\Psi$ represent the same fields as in the
previous sections, and the convention for the crossed out forms of
rank $k$ is that they are contracted with $k$ gamma matrices with a
normalization factor $\frac{1}{k!}$. Then, we have the redefinition for the scalar fields: 
\be \Phi \equiv -\scal{\phi_1-\phi_2} \hspace{2cm} \bar\Phi\equiv
-\frac{1}{2}\scal{\phi_1+\phi_2} \label{twist1} \ee
The twisted action that one obtains in this way is 
\begin{multline}
\label{93}
\int_M d^8x\trace\biggl(-\frac{1}{4}F_{\mu\nu}F^{\mu\nu}+\eta D_\mu\Psi^\mu +
D_\mu \bar\Phi D^\mu\Phi + 4\chi^{\mu\nu}D_\mu\Psi_\nu \biggr.\\*\biggl.
+2\bar\Phi\Psi_\mu\Psi^\mu + 2\Phi
\chi_{\mu\nu}\chi^{\mu\nu} + \Phi \eta^2 + \frac{1}{2}[\Phi,\bar\Phi]^2 \biggr)
\end{multline}
It is the same action as that obtained in section \ref{action}, from
the demand of $s$ and $\delta$ invariances, after the 
integration of the auxiliary field $T$, modulo some rescalings, and
up to the sum of a 
topological term:
\be -\frac{1}{2} \int_M C_{\,\wedge}\trace\scal{F_{\, \wedge} F} \ee

By using the decomposition by twist of the spinorial
supersymmetry parameter $\epsilon = \scal{\theta + \Theta^{\mu\nu}
 \gamma_{\mu\nu} + i \vartheta^\mu \gamma_\mu } \zeta $, one gets
twisted generators $Q,Q_\mu,Q_{\mu\nu}$. 
($Q$ and $\kappa^\mu Q _\mu$ are truly identified with 
 the BRST operators $s_c$ and $\deltac$, in their equivariant
form, both in 8 and 4 dimensions.)

Both charges
$\scal{\overline{\zeta} \mathcal{Q}}$ and $i \scal{\overline{\zeta} \ba \kappa
 \mathcal{Q}}$ completely constrain the supersymmetric theory. In
this sense, 
the number of relevant supersymmetries can be reduced 
to five real supercharges 
in four dimensions (as already
observed in 
 \cite{Tet}) and to nine real supercharges in
in eight dimensions. 

It is very instructive that this reduced number of relevant generators
can be directly constructed from one extended horizontality
condition, defined in the Yang--Mills principal bundle.

As a further remark,
 the tensor generator of supersymmetry
cannot be closed 
off-shell in eight dimension \cite{glouk}, 
contrarily to the case of 4 dimensions
\footnote{From a technical point of view, the difference between four
 and eight dimensions amounts the fact that one can construct an
antiselfdual $2$-form as a bilinear combination of two other in four
dimensions, by the use of
\be 
{{P^-}_{\mu\nu}}^{\theta\tau}
{{P^-}_{\theta}}^{\eta\sigma\rho}{{P^-}_{\tau\eta}}^{\kappa\lambda}
\ee
but not in eight, because this term is zero in this case. }. It is
unknown if a system of 
auxiliary fields can be introduced to close the algebra of maximal
supersymmetry. The existence of the tensor
symmetry is not foreseen from the point of view of TQFT's, and its
existence seems unnecessary, since the $Q$ and $Q_\mu$ symmetries are
enough to determine the supersymmetry action.


\section{Equivariant Topological Field Theories}

The observables of the topological theories that we discussed so far 
are defined as classes of the ordinary de Rham cohomology
of the extended exterior derivative
$\t{d} = d+s$ acting on 
$M \times \t{\mathcal{B}}^\ast$,
where $M$ is the manifold on which the topological theory is formulated
and $\t{\mathcal{B}}^\ast$ is the space of
gauge orbits of irreducible framed connections \cite{Mo,BaSi}.
In these theories, the scalar ghost-for-ghost field $\Phi$
goes to zero at infinity,
or in other words, has vanishing vacuum expectation value. 
Since, as we can see from (\ref{twist1}), 
the $\Phi$ field is related to the scalar fields of the 
supersymmetric Yang--Mills theory,
it is interesting to construct a topological theory
whose scalar fields acquire a non-vanishing vacuum expectation value
\cite{vev}. This topological field theory can be obtained by 
considering the 
equivariant extension of the construction of \cite{BaSi}
with respect to the Lie algebra ${\mathfrak{h}}$ of an Abelian
subgroup $H\subset G$ 
acting on $ \t{\mathcal{B}}^\ast$.
For example, for $G=SU(N)$ one can consider the maximal 
Abelian subgroup $H=U(1)^{N-1}$, which is the suitable choice
for the Seiberg--Witten model. 
Moreover, we also consider an equivariant extension with respect
to a compact Abelian group of isometries $K$ of $M$. This corresponds to a kind
of spontaneous breaking of the symmetries of $M$
down to $K$. In fact,
as we will see in the following, the resulting equivariant topological theory
corresponds to the twisted version of the Super Yang--Mills theory
on a non--trivial gravitational background.

The equivariant formulation allows for the use of 
a powerful localization formula \cite{berline}
that reduce the integral over the equivariant forms on $M \times
\t{\mathcal{B}}^\ast$ 
to a sum over the isolated fixed points of the 
$K\times H$ symmetry\footnote{Notice
that the fixed points are isolated only in this case:
considering the equivariance with respect to only one of the two
groups, $K$ or $H$, is not enough to localize to isolated points.}.
The results on
the ordinary cohomology may in general 
be recovered by sending to zero the parameters
associated to this symmetry with a suitable prescription.
In this sense, the equivariant extension can be thought as a
very useful regularization procedure for the topological invariants. 
The localization formula 
has been extended for supermanifolds in \cite{equiv}
and exploited in the four dimensional 
case to compute the integral on the instanton moduli space,
recovering the non--perturbative contribution to the low--energy
Seiberg--Witten effective action \cite{equiv,nekrasov,flume}.

In the following we will discuss the equivariant 
extension of our horizontality conditions and obtain from them
the scalar and vector topological symmetries along the same
lines of the previous sections. Then we will untwist our
topological theory and show its relationship with the
supersymmetric theory on the $\Omega$-background introduced
in \cite{nikita}. 

\subsection{Equivariant horizontality condition}

Let us define the Weil complex corresponding to the equivariant
cohomology $$ H^{\bullet}_{K\times H} \scal{ M \times
 \t{\mathcal{B}}^\ast }$$
equivariant with respect to the action of $K\times H$ on
$\t{\mathcal{B}}^\ast$. 
The action of $K$ on $\t{\mathcal{B}}^\ast$ can be defined as follows.
The action of $K$ on $M$
can be lifted to an action on the principal bundle $P$ by the
use of a background connexion $\bA$. This action induces a pullback
action on the equivariant forms on $P$, which defines the action of $K$
on $\t{\mathcal{B}}^\ast$. 
The relevant equivariant differential
$s_{\mathfrak{k}}$ on the Weil complex is defined as usual on an
equivariant form $w(\xi)$ by 
\be
 \scal{ s_{\mathfrak{k}} w } (\xi) \equiv \scal{ s + I_{\xi^*}}
 w(\xi) 
\label{s-e}
\ee
where $\xi^*$ is the vector field of $T P$ generating the action on
$\t{\mathcal{B}}^\ast$ associated to $\xi = ( \Omega , a ) \in
\mathfrak{k} \oplus \mathfrak{h}$. $\xi^*$ decomposes into the
horizontal lift $\u^h$ of the vector field $\u$ of $TM$ generated by
the element $\Omega$ of $\mathfrak{k}$ and the fundamental vector
$a^v$ associated to the element $a$ of $\mathfrak{h}$,
\be \xi^* \equiv (\u^h,a^v) \in TH \oplus TV \ee
 The closure of the equivariant BRST operator
(\ref{s-e}) reads on a generic form 
\be s_{\mathfrak{k}}^2 = \L_{\xi^*} \ee
so that $s_{\mathfrak{k}}$ is a nilpotent operator
on equivariant forms. This is locally expressed on $M$ 
\be s_{\mathfrak{k}}^2 = \L_{\u} + \delta_{\mathrm{gauge}} ( i_{\u}
\bA - a ) \ee
The explicit representation of the operator (\ref{s-e})
on anti--self--dual gauge connections
has been discussed in detail in \cite{flume,equiv,super}.
Since $i_{\xi^*}$ commutes with $s_{\mathfrak{k}}$,
we have the following nilpotent operator on the whole complex
(not restricted to its invariant subcomplex)
\be ( d + s_{\mathfrak{k}} - i_{\u^h + a^v} )^2 = 0 \ee
So we can define $s_{\mathfrak{k}}$ as usual by the use of an
horizontality condition
\be\label{honil}
(d + s_{\mathfrak{k}} - i_{\u^h + a^v} ) \scal{A + c} + \scal{A+c}^2 =
F + \Psi + \Phi 
\ee
and its associated Bianchi identity. Moving the term $i_{\xi^*} A$ to
the right hand side we obtain the definition of the equivariant
curvature \cite{berline}
\be
(d + s_{\mathfrak{k}}) \scal{A + c} + \scal{A+c}^2 = F + \Psi +
\Phi_{\mathfrak{k}} 
\label{equi}
\ee 
Notice that on the right hand side of (\ref{equi}) it appears
the equivariant extension of the scalar field 
$\Phi_{\mathfrak{k}}=\Phi_{\mathfrak{h}} + i_\u (A-\bA)= \Phi + a +
i_\u (A-\bA)$, 
where $\mu({\xi})= a + i_\u (A-\bA)$
is the moment of the vector field ${\xi^*}$ \cite{berline}.
This deformation of the scalar field is precisely that 
found in the explicit computations 
on the instanton moduli space in four dimensions 
\cite{matone}.
The field $\Phi_{\mathfrak{h}}$
has a non--trivial vacuum expectation value
in the Cartan subalgebra of the group $G$, due to the presence
of the term $a$ \cite{flume,equiv,matone}.
Notice that the vector $\xi^*$ has ghost number $2$.

The dual version of equation (\ref{honil}) is naturally
defined with the use of another Killing vector $\bu$ on $M$ 
and another element $\bar a$ of $\mathfrak{h}$ which define a vector
$\bar \xi^* = \bu^h + \bar a^v $ on $P$, all with ghost number $-2$,
and reads 
\be
(d + \delta_{\mathfrak{k}} - |\kappa|^2 i_{\bu^h + \bar a^v}) \scal{A
 + |\kappa| \bar c} + 
\scal{A + |\kappa| \bar c}^2 = F + g(\kappa) \eta + i_\kappa \chi +
|\kappa|^2 \bar\Phi 
\label{equi1}
\ee

Exactly in the same way as in the previous sections, we can combine
the horizontality conditions (\ref{honil}) and (\ref{equi1}) 
into a single one
which will define the equivariant BRST operator as well as the
corresponding vector symmetry. The extended horizontality condition is 
\begin{gather}
( d + s_{\mathfrak{k}} + \delta_{\mathfrak{k}} 
- i_{\kappa^{h} + \xi^* + |\kappa|^2 \bar \xi^*}) \scal{ A
 + c + |\kappa| \bar c} + \scal{ 
 A + c + |\kappa| \bar c}^2 \hspace{40mm}\CR
\hspace{50mm} = F + \Psi + g(\kappa) \eta + i_{\kappa^{h}}
\chi + \Phi + |\kappa|^2 \bar\Phi
\label{eH}
\end{gather}
and its associated Bianchi relation
\begin{gather}
( d + s_{\mathfrak{k}} + \delta_{\mathfrak{k}} 
- i_{\kappa^{h} + \xi^* + {| \kappa|^2} \bar \xi^*}) \scal{F +
 \Psi + g(\kappa) 
 \eta + i_{\kappa^{h}} \chi + \Phi + |\kappa|^2 \bar\Phi}
\hspace{30mm}\CR \hspace{25mm} + \,[
A + c + |\kappa| \bar c\,,\, F + \Psi + g(\kappa) \eta +
i_{\kappa^{h}} \chi + \Phi + |\kappa|^2 \bar\Phi ] = 0
\label{eB}
\end{gather}
By following the same procedure described in the previous sections
for the ordinary topological field theory, 
and redefining now
\be
\label{delta3}
s_c \equiv s_{\mathfrak{k}} + \delta_{\mathrm{gauge}}(c) \hspace{10mm} 
\deltac \equiv
\delta_{\mathfrak{k}} + \delta_{\mathrm{gauge}}(|\kappa| \bar c)
\ee
we can extract from
(\ref{eH}) and (\ref{eB})
the complete transformations of the fields
\be
\label{odelta}
\begin{array}{rclcrcl}
s_c A &=& \Psi &\hspace{10mm}& \deltac A &=& g(\kappa)
\eta + i_\kappa \chi \\* 
s_c \Psi &=& -d_A \Phi +i_{\u} F & & \deltac \Psi
&=& i_\kappa \scal{ T+ F} + g(\kappa) s_c \eta
\\* 
s_c \Phi &=& i_{\u} \Psi & & \deltac \Phi &=& 
i_\kappa \Psi + (\kappa\cdot \u) \eta + i_{\u} i_\kappa \chi \\* 
&&&&&&\\*
s_c \bar\Phi &=& \eta + i_{\bu} \Psi & & \deltac \bar\Phi
&=& (\bu\cdot \kappa) \eta + i_{\bu} i_\kappa \chi \\* 
s_c \eta &=& [\Phi, \bar\Phi] +\Lc_{\u}
 \bar\Phi - \Lc_{\bu} \Phi + i_{\bu} i_{\u} F & & \deltac \eta &=&
\Lc_\kappa \bar\Phi - i_\kappa i_{\bu} F \\* 
&&&&&&\\*
s_c \chi &=& T & & \deltac \chi
&=& \frac{n}{2} \scal{g(\kappa) (d_A \bar\Phi - i_{\bu} F)}^- \\* 
s_c T &=& [\Phi, \chi] + \Lc_{\u} \chi & & \deltac T &=& -\frac{n}{2}
\scal{g(\kappa) (d_A \eta + [\bar\Phi, \Psi] + \Lc_{\bu} \Psi)}^- + \Lc_\kappa
\chi \\*
\end{array}
\ee
We remark that in the four dimensional case
the $s_c$ transformations on the fields in the first column 
of (\ref{odelta}) induce exactly the BRST transformations on the instanton
moduli space that have been used for the localization in 
\cite{nekrasov,flume,equiv}. The transformations in the Faddeev--Popov
sector read 
\be
\label{odelta1}
\begin{array}{rclcrcl}
s_{\mathfrak{k}} \,c &=& \Phi+ a - c^2 +i_{\u} \scal{A-\bA} &\hspace{1cm} &
\delta_{\mathfrak{k}}\, c &=& i_\kappa 
\scal{A-\bA} - |\kappa| b\\* 
s_{\mathfrak{k}} \, \bar c &=& b - [c, \bar c]& & \delta_{\mathfrak{k}}
\, \bar c &=& |\kappa| \scal{ \bar\Phi + \bar a - {\bar c}^2 + i_{\bu}
 \scal{A - \bA}} \\* 
s_{\mathfrak{k}} \, b &=& [\Phi, \bar c] - \Lc_{\u} \bar c - [c,b] & &
\delta_{\mathfrak{k}} \, b &=& |\kappa| \scal{ \Lc_\kappa \bar c - \eta}
\end{array}
\ee
The algebra (\ref{odelta}), (\ref{odelta1})
closes off-shell, provided that $[\u, \bar\u]=\L_{\u}\bar\u=0$,
$ d(\kappa\cdot\bu) = 0$. Moreover, $d g(\u)$ and $d g(\bu)$ must be
selfduals in the eight dimensional case. Then, one has, but on the
Faddeev-Popov sector:
\begin{gather}
s_c^2 = \delta_\mathrm{gauge} (\Phi+ i_{\u} A) + \L_{\u}
\hspace{10mm} \deltac^2 = \delta_\mathrm{gauge} (\bar\Phi+ i_{\bu} A)
+ \L_{\bu} \CR \{ s_c, \deltac\} = \L_\kappa +
\delta_\mathrm{gauge} (i_\kappa A) \label{squaresd}
\end{gather}
The $s_{\mathfrak{k}}$ and $\delta_{\mathfrak{k}}$ symmetries 
completely constrain the classical 
action also in the equivariant case. The details of this
computation are given in appendix \ref{gf}. 
The action of the equivariant topological theory is
$s_{\mathfrak{k}}\, \delta_{\mathfrak{k}}$-exact
\be
S = s_{\mathfrak{k}} \, \delta_{\mathfrak{k}} \int_M \usk \, \mathscr{F} \ee
with
\be \label{sdelta}
\mathscr{F} = \trace \Biggl( \frac{1}{2} 
 g(\kappa)_{\,\wedge}C_{\,\wedge}\Scal{ \scal{ A- \bA}_{\,
 \wedge}\scal{ F + \bF} - \frac{1}{3} \scal{A- \bA}^3}
+ \scal{g(\kappa) \eta + {i_\kappa \chi}}_{\, \wedge}
 \star \Psi \Biggr) 
\ee 
and still displays an intriguing relationship with the 
Chern--Simons action functional.
By acting with $\delta_{\mathfrak{k}}$ in
(\ref{sdelta}) one gets
\begin{multline}
I= 
s_{\mathfrak{k}} \int_M\trace \Biggl(\chi\star(F+\frac{2}{n} T) + \Psi \star \Scal{
d_A \bar\Phi -i_{\bu} F} \Biggr .\\*
+ \star\eta \Scal{ [\Phi,\bar\Phi] + \Lc_{\u} \bar\Phi -\Lc_{\bu} \Phi +
 i_{\bu} i_{\u} F} \Biggr) \label{gaom}
\end{multline} 
and finally by acting with $s_{\mathfrak{k}}$
\begin{multline}
I= \int_M \trace \Biggl( F^-\star F^- - \Scal{d_A \Phi - i_{\u} F}
\star \Scal{d_A \bar\Phi - i_{\bu} F} + \chi \star d_A \Psi - \Psi \star d_A
\eta \Biggr . \\*
 + T \star \scal{F+ \frac{2}{n} T} - \frac{2}{n} \chi \star \Scal{
 [\Phi, \chi] + \Lc_{\u} \chi} - \eta \star
 \Scal{[\Phi, \eta] + \Lc_{\u} \eta } \\*- \Psi \star \Scal{ [\bar\Phi, \Psi] +
 \Lc_{\bu} \Psi} + \Scal{ [\Phi, \bar\Phi] + \Lc_{\u} \bar\Phi -
 \Lc_{\bu} \Phi + i_{\bu} i_{\u} F }^2 \Biggr)
\label{aho}
\end{multline}

By comparing the equivariant topological action
(\ref{aho}) to the topological action (\ref{93}) one get a simple
rule to pass from one to the other. In the case discussed 
in Sect.4 the $s_c$ and $\delta_{\bar c}$ operators are nilpotent modulo
gauge transformations, {\i.e.} $s_c^2=\delta_{\rm gauge}(\Phi)$ and
$\delta_{\bar c}^2=\delta_{\rm gauge}(\bar\Phi)$.
In the equivariant case, the nilpotency is also verified modulo
reparametrizations
along the Killing vectors $\u$ and $\bu$, as one can see from the first line
of Eq.~(\ref{squaresd}). To pass from the ordinary topological theory to the 
equivariant one, we have to make the substitution
\be
\delta_{\rm gauge}(\Phi) \rightarrow \delta_{\rm gauge}(\Phi + i_\u A) + 
\L_{\u} \quad\quad\quad 
\delta_{\rm gauge}(\bar\Phi) \rightarrow 
\delta_{\rm gauge}(\bar\Phi + i_{\bar\u} A) + 
\L_{\bu}
\ee
This amounts to the redefinitions 
\begin{gather}
d_A \Phi \rightarrow d_A \Phi - i_\u F \quad\quad\quad 
d_A \Phi \rightarrow d_A \bar\Phi - i_{\bar\u} F \CR
\quad [\Phi,\bar\Phi] \rightarrow [\Phi,\bar\Phi] 
+ \Lc_{\u} \bar\Phi - \Lc_{\bu}\Phi + i_{\bu} i_\u F
\label{r1}
\end{gather}
for the bosonic fields, and
\begin{gather}
 [\Phi,\chi]\rightarrow [\Phi,\chi]+\Lc_\u\chi
\quad\quad
 [\Phi,\eta]\rightarrow [\Phi,\eta]+\Lc_\u\eta
\CR
\quad\quad \ [\bar\Phi,\Psi]\rightarrow [\bar\Phi,\Psi]+\Lc_{\bu}\Psi
\label{r2}
\end{gather}
for the fermion fields. 
One can check that, by doing the substitutions
(\ref{r1}) and (\ref{r2}) in the topological action (\ref{93}),
one obtains the equivariant topological action (\ref{aho}).

Finally,
we see that in the equivariant topological theory,
the scalar field $\Phi_{\mathfrak{h}}$ has a non--trivial expectation value.
In the following subsection we will show that the equivariant action
(\ref{aho}) can be related by twist to the Super Yang--Mills
theory on the $\Omega$-background introduced in \cite{nikita}.

\subsection{Dimensional reduction and $\Omega$ background}

The so-called $\Omega$-background can be introduced by considering
a non-trivial dimensional reduction of the Super Yang--Mills theory
on a torus.
In this dimensional reduction, the original theory is defined on a 
Riemannian fiber bundle $E$
\bea 
M \rightarrow & E & \CR
 & \downarrow& \CR
 & T^{m-n}& 
\eea
such that 
the manifold $M$ on which the dimensionally
reduced theory lives is fibered on the torus.
Eventually, we can define the metric as
follows : 
\be G \equiv \delta_{\alpha\beta} \, dy^\alpha \otimes dy^\beta +
g_{\mu\nu} \scal{dx^\mu + \u_\alpha^\mu dy^\alpha }\otimes
\scal{dx^\nu +\u_\beta^\nu dy^\beta} \label{G}
\ee
Here, the $x^\mu$ are local coordinates on $M$, $y^\alpha$ are coordinates
on $T^{m-n}$ and $\u_\alpha$ are vector fields on $M$. 
It is natural to require that the metric $G$ does not
depend on the torus coordinates; in fact, any non--trivial dependence
 would forbid a consistent cancellation of non-zero modes
in the limit of zero radius.

In order to have a supersymmetric theory, one requires
the existence of, at least, one supersymmetry generator,
and thus the existence of a covariantly constant spinor field on the
manifold. This implies that both manifolds $M$ and $E$ are 
Ricci flat \cite{Joyce} :
\be R^{\hspace{-5mm} E}_{mn} = 0 \hspace{1cm} R^{\hspace{-5.5mm}
 M}_{\mu\nu} = 0 \label{pri}\ee
These equations turn into constraints
on the vector fields $\u_\alpha$ 
\bea
\L_{\u_\alpha} g_{\mu\nu} &=& 0 \CR
\left[ \upsilon_\alpha ,\upsilon_\beta \right] &=& 0 \CR
d \star d g(\u_\alpha) &=& 0 \label{homo}
\eea
The first equation in (\ref{homo}) implies that $\u_\alpha$'s
 are Killing vectors for the manifold $M$, while the second
imposes that they commute, {\it i.e.} 
$\L_{\u_\alpha} \u_\beta = 0$.
The vectors $\u_\alpha$ can be mapped upon the Killing vectors that we
used in the construction of the equivariant topological field theory
in the previous subsection.
The last equation in (\ref{homo})
imposes further restrictions on the $\u_\alpha$'s
in order to preserve the supersymmetry. 
These conditions are not present in the topological theory.
In fact, as we will see in detail in the following, they
can be relaxed at the price of breaking the $SO(n)$ rotation
invariance of the supersymmetric theory to the special holonomy subgroup
required to define the topologically twisted theory.

Let us now work out the case of the eight-dimensional 
super Yang--Mills theory.

\subsubsection{Supersymmetric formulation in eight dimensions}

The ten-dimensional vielbeins corresponding to the metric (\ref{G}) are
\be
\label{viel}
e^A_m \hat{=} \left (\begin{array}{cc} 
\hspace{3mm} e^a_\mu \hspace{3mm}& \,i_{\u_\beta} e^a \,\\* 0 & \delta^\alpha_\beta 
\end{array} \right ) 
\hspace{1cm}
e^m_A \hat{=} \left (\begin{array}{cc} 
\hspace{3mm} e_a^\mu \hspace{3mm}& \hspace{3mm}0 \hspace{3mm} \\* - \u^\mu_\alpha & \delta_\alpha^\beta 
\end{array} \right ) 
\ee
The dimensional reduction of the ten-dimensional Yang--Mills curvature
reads
\be
 \frac{1}{2} F_{mn}\, {dx^m}_\wedge dx^n = \frac{1}{2} F_{\mu\nu}\,
{dx^\mu}_\wedge dx^\nu + D_\mu \phi_\alpha \,{dx^\mu}_\wedge dy^\alpha +
\frac{1}{2} [\phi_\alpha, \phi_\beta] \,{dy^\alpha}_\wedge dy^\beta \CR
\ee
We can write this curvature in locally flat coordinates
by using the vielbeins (\ref{viel})
\begin{multline}
\label{granf}
 \frac{1}{2} F_{AB}\, {e^A}_{\wedge} e^B = \frac{1}{2} F_{ab}
\,{e^a}_\wedge e^b + \Scal{e_a^\mu D_\mu \phi_\beta - 
 \u_\beta^\nu e^\mu_a F_{\mu\nu}} \,{e^a}_\wedge dy^\alpha \\*
 \,\, + \frac{1}{2} \Scal{[\phi_\alpha, \phi_\beta] -
 \u^\mu_\alpha D_\mu \phi_\beta + \u^\mu_\beta D_\mu \phi_\alpha +
 \u^\mu_\alpha \u^\nu_\beta F_{\mu\nu}}\, {dy^\alpha}_\wedge dy^\beta
\end{multline}
By plugging (\ref{granf}) into the action
\be
\int_M d^8x \sqrt{g} \, \trace \Biggl( -\frac{1}{4} F_{AB} F^{AB} +
\frac{i}{2} \scal{\, \overline{\Lambda} \, \Gamma^A D_A \, \Lambda\, } \Biggr)
\ee
one can read the bosonic part of the eight-dimensional action.
Concerning the fermionic part, we have 
to decompose the contraction of the covariant derivative $e^m_A D_m$.
To simplify the computation we observe that 
\be 
e^m_A D_m = \Lc_{e^m_A}
\ee
since $e^m_A$ is covariantly constant. The Lie
derivative is independent of the Riemannian connection and thanks to this
property, one has :
\be 
e^m_A D_m \hat{=} \Scal{\, e_a^\mu D_\mu\,\, ,\, \, \phi_\alpha-
 \Lc_{\u_\alpha} \, } 
\ee
The ten-dimensional gamma matrices are related to those in eight dimensions 
as follows: 
\be
\Gamma^m \equiv \sigma_2 \otimes \gamma^\mu , \sigma_2 \otimes
\gamma_9 , \sigma_1 \otimes 1
\ee
Using the above equations one obtains the action and its
supersymmetries on a eight dimensional pseudo-Riemannian 
manifold. By 
extending to eight dimensions the Wick rotation on the fermions, which is defined in
\cite{wick}, one can Wick-rotate this theory 
to a Riemannian manifold, as follows :
$$x^0 \rightarrow e^{-i\theta} x^0 \hspace{2cm}
\gamma^8 \equiv i \gamma^0 $$
$$A_\mu \rightarrow (e^{i\theta}A_8,A_i) $$
\be
\begin{array}{rclcrcl}\medskip
\lambda &\rightarrow& e^{\frac{1}{2}\theta\gamma^8\gamma_9}\lambda
&\hspace{2cm} & \lambda^\dagger &\rightarrow &\lambda^\dagger
e^{\frac{1}{2}\theta\gamma^8\gamma_9} \\*\medskip
\phi_1 &\rightarrow& e^{i\theta} \phi_1 & &
\phi_2 &\rightarrow &\phi_2 \\*\medskip
\u_1 &\rightarrow& e^{i\theta} \u_1 & &
\u_2 &\rightarrow &\u_2 
\end{array}
\ee
Eventually, one sets $\theta=\frac{\pi}{2}$, and one gets 
the Euclidean action 
\begin{multline}\label{nikita}
\int_M d^8x \sqrt{g} \,\trace\Biggl( -\frac{1}{4} F_{\mu\nu}F^{\mu\nu} -
\frac{1}{2}
\Scal{ D_\mu \phi_\alpha + \u^\nu_\alpha F_{\nu\mu}}\Scal{D^\mu
 \Phi^\alpha + \u_\sigma^\alpha F^{\sigma\mu}} +\frac{i}{2} \left(\overline{\lambda}
 \baa D\lambda\right) \biggr.\\*
 -\frac{1}{2}\Scal{[\phi_1,\phi_2] -\mathscr{L}_{\u_1} \phi_2 +
 \mathscr{L}_{\u_2} \phi_1 + \u_1^\mu \u_2^\nu F_{\mu\nu}}^2 \\*
 +\frac{i}{2}\Scal{\overline{\lambda}\gamma_9\scal{[\phi_1,\lambda]-\mathscr{L}_{\u_1}\lambda}}
 +\frac{1}{2}\Scal{\overline{\lambda}\scal{[\phi_2,\lambda]-\mathscr{L}_{\u_2}\lambda}} \Biggr)
\end{multline}
We recall that the spinor Lie derivative is defined by
\be
\mathscr{L}_{\u}\lambda = \u^\mu D_\mu \lambda + \frac{1}{2}
D_\mu\u_\nu \gamma^{\mu\nu}\lambda
\label{spinor-lie}
\ee
where $\gamma^{\mu\nu} \equiv \frac{1}{4} [\gamma^\mu, \gamma^\nu]$.

The action (\ref{nikita}) has the following supersymmetry :
\bea\label{gggg} &\delta A_\mu &= -i(\overline{\epsilon} \gamma_\mu \lambda)\CR
 &\delta \phi_1 &= -(\overline{\epsilon} \gamma_9 \lambda) - i
 \scal{\overline{\epsilon} \ba \u_1 \lambda}\CR
 &\delta \phi_2 &= -(\overline{\epsilon} \lambda) -
 i\scal{\overline{\epsilon} \ba \u_2 \lambda}\CR
 &\delta \lambda &= \ba F\epsilon-i\gamma_9 \Scal{\baa D\phi_1 +
 \hspace{3mm} \diagup \hspace{-7mm} i_{\u_1} F} \epsilon
-i\Scal{\baa D\phi_2 + \hspace{3mm}\diagup \hspace{-7mm} i_{\u_2} F}\epsilon \CR
& & \hspace{10mm} +\gamma_9 \Scal{[\phi_1,\phi_2] -
 \mathscr{L}_{\u_1} \phi_2 + \mathscr{L}_{\u_2} \phi_2 + i_{\u_2}
 i_{\u_1} F}\epsilon \CR \label{omesup}
\eea
In fact, the symmetry holds only if the covariantly constant spinor
$\epsilon$ is constant along the flow of the $\u_\alpha$'s. 
Thus $\Omega^\alpha_{ab} \equiv \frac{1}{2} \scal{d
 g(\u^\alpha)}_{ab}$
must define a degenerated matrix $\ba \Omega^\alpha$. 
In eight dimensions, a constant spinor only exists
on a Joyce manifold, and its 
Lie derivative can be written as 
\be \L_{\u} \varepsilon_{\alpha} = \frac{1}{2} {\ba
\Omega_\alpha}^\beta \varepsilon_{\beta} 
\ee
which is zero if and only if $\Omega$ is selfdual.
Thus the constant spinor remains invariant under 
the isometries generated by $\u_\alpha$'s 
only if the matrix $\Omega_\alpha$ is
selfdual. Such vectors verify the third equations in (\ref{homo}) 
$d\star d g(\u_\alpha)$=0. 
The explicit form of the action (\ref{nikita}) for $M \cong \mathds{R}^8$
is displayed in Appendix C.

\subsubsection{The twisted theory for $\Omega$ backgrounds}
\label{omegatwist}

The modifications of the supersymmetric theory are formally quite mild when one
introduces the $\Omega$ backgrounds. They are completely determined at
the purely bosonic level. Thus, all twist operations must remain
identical, and one can define the twisted scalar and vector 
operators $Q$ and $Q_\mu$ from Eq.(\ref{gggg}).
 To compute the twisted version of the action (\ref{nikita}), we define: 
\be
\u = \u_1 - \u_2 \hspace{1cm} \bu = \frac{1}{2} \scal{\u_1 + \u_2}
\label{twist2} 
\ee
We obtain :
\begin{multline}
I= \int_M \trace \Biggl( -\frac{1}{2} F\star F - \Scal{d_A \Phi - i_{\u} F}
\star \Scal{d_A \bar\Phi - i_{\bu} F} -4 \chi \star d_A \Psi - \Psi \star d_A
\eta \Biggr . \\*
- 2 \chi \star \Scal{ [\Phi, \chi] + \Lc_{\u} \chi} - \frac{1}{2} \eta \star
 \Scal{[\Phi, \eta] + \Lc_{\u} \eta } \\*- \Psi \star \Scal{ [\bar\Phi, \Psi] +
 \Lc_{\bu} \Psi} + \frac{1}{2} \Scal{ [\Phi, \bar\Phi] + \Lc_{\u} \bar\Phi -
 \Lc_{\bu} \Phi + i_{\bu} i_{\u} F }^2 + 2 T\star T \Biggr)
\label{aho-tw}
\end{multline}
One can verify that this action is $Q$ exact, up to a topological
term and can be written as follows: 
\begin{multline}
I= 
-\frac{1}{2}\int_M\trace\scal{\,\,C_{\,\,\wedge}\, F_{\,\,\wedge}\, 
F\,\,}\\*
+ Q \int_M\trace \Biggl(-2\chi\star(F-T) + \Psi \star \Scal{
d_A \bar\Phi -i_{\bu} F} \Biggr .\\*
+ \frac{1}{2}\star\eta \Scal{ [\Phi,\bar\Phi] + \Lc_{\u} \bar\Phi -\Lc_{\bu} \Phi +
 i_{\bu} i_{\u} F} \Biggr)
\end{multline} 
We thus recover by twisting the action of the equivariant topological 
theory (\ref{aho})\footnote{This hold true, up to a
rescaling in (\ref{gaom}) which leaves the BRST operator
invariant. That is of a factor $-2$ for $\chi$ and $T$, a factor
$\frac{1}{2}$ for $ \bar\Phi, \eta $ and $\bu$, and a global factor
$2$ on the action. We must also add the substitution $T \rightarrow T
+ F^-$.}, and discover that the scalar and vector topological 
symmetries defined from the equivariant horizontality conditions 
correspond to the twisted supersymmetries of the Super Yang--Mills
theory on the $\Omega$-background.
Notice that the action in the 
twisted formulation (\ref{aho-tw}) is BRST--exact, and so
BRST closed, for all commuting vector fields $\u_\alpha$,
so that the matrices $\Omega^\alpha_{ab}$ are no longer required 
to be self--dual in order to have a BRST--closed action.
This can be understood as follows: if we consider
the twisted theory for a generic $\Omega$-background 
({\it i.e.} $\Omega^\alpha_{ab}$ generic matrices)
and we untwist it, we get a term
\be
\frac{1}{4} \lambda^\alpha {\ba \Omega^+_\alpha}^\beta \lambda^\beta -
\frac{1}{2} \lambda_{\aalpha} \scal{{\ba \overline{\Omega}^{+\, \aalpha}}_{\bbeta}-{\ba \overline{\Omega}^{\, -\, \aalpha}}_{\bbeta}}
\lambda^{\bbeta} 
\label{break}
\ee
where the plus and the minus stand for the self--dual and anti--self--dual
projections with respect to the Cayley four form $C$.
The term (\ref{break}) 
breaks the rotation invariance from
$SO(8)$ to $Spin(7)$, effectively twisting the theory.
 

\section{Conclusion}

On manifolds of reducible tangent bundle, the existence of a
covariantly constant vector field allows one to extend the
horizontality condition. This extension define two nilpotent
topological operators, 
the usual scalar one, and the vector topological operator. These two
operators define 
a closed off-shell algebra, in a globally well-defined way. In order
to make contact with known results, we observe that the 
dimensional reduction on a circle with tangent vector $\kappa$ of this
horizontality condition is nothing but the BRST-antiBRST
horizontality condition, which also defines two topological 
charges of a resulting balanced topological theory, as it was defined
in \cite{moore}. The 
consistency of the algebra needs the existence of the concept of
selfduality or antiselfduality. In eight dimensions, this implies
that the manifold is of a Joyce type. 
 The use of the vector symmetry permits one to raise the
 indetermination of the topological 
gauge function, and eventually of a topological BRSTQFT action that
determines by twist supersymmetry. 

The invariance of the action under the vector symmetry is in fact equivalent
to the conservation of its Noether current, which turns out to be a
BRST--antecedent of the energy momentum tensor. A more conventional
construction would be the definition of a BRSTQFT from the last
condition, but it would obscure the geometrical interpretation. 

 
This algebraic construction of topological theory extends to the case
of $\Omega$ background. The extended differential is understood as the
equivariant differential with respect to the action of an isometry
group of the physical space, and the observables of the theory are in
the equivariant cohomology of this differential. 

Beyond the mathematical interpretation of the fields occurring in
BRSTQFT, it is striking that the extended horizontality condition
also provides a geometrical 
construction of a subalgebra of (possibly maximal) supersymmetry which
can is closed off-shell and {\it completely} determines the
action. Thus, it determines 
 the whole supersymmetric algebra in the flat space limit, and the
 question of having no finite set of auxiliary field for the
 superPoincar\'e algebra becomes irrelevant. These results are
 compatible with dimensional reduction, and apply therefore to
 different cases of supersymmetry, in other dimensions.


\appendix

\section{Algebra of the octonionic 4-form}
\label{a3}
We define the two projectors :
\bea
{{P^-}_{\mu\nu}}^{\sigma\rho} &\equiv& \frac{1}{4}
\left(\delta_{\mu\nu}^{\sigma\rho}-\frac{1}{2}{C_{\mu\nu}}^{\sigma\rho}\right)
 \CR
{{P^+}_{\mu\nu}}^{\sigma\rho} &\equiv& \frac{3}{4}
\left(\delta_{\mu\nu}^{\sigma\rho}+\frac{1}{6}{C_{\mu\nu}}^{\sigma\rho}\right)
 \CR
\eea
and from the formula \cite{aB}
\bea C_{\mu\nu\sigma\rho}C^{\kappa\lambda\theta\rho} =
6\delta_{\mu\nu\sigma}^{\kappa\lambda\theta} - 9
{C_{[\mu\nu}}^{[\kappa\lambda}\delta_{\sigma]}^{\theta]}\eea
we can show that
\bea
{{P^-}_{\sigma\rho\{\mu}}^{\theta}{{P^-}_{\nu\}\theta}}^{\kappa\lambda}
&=& \frac{1}{8} \delta_{\mu\nu}{{P^-}_{\sigma\rho}}^{\kappa\lambda} \CR
\left(\delta_{\mu\nu}^{\{\theta|\tau}+\frac{1}{2}{C_{\mu\nu}}^{\{\theta|\tau}\right){{P^-}^{\sigma\}\rho\kappa}}_{\tau}
&=& -\delta^{\theta[\sigma}{{P^-}_{\mu\nu}}^{\rho]\kappa} \CR
\Scal{\delta^{[\mu|\theta}_{\sigma\rho} + \frac{1}{2}
 {C_{\sigma\rho}}^{[\mu|\theta}}{P^{-\,
 \kappa\lambda|\nu]}}_\theta &=& \frac{1}{4} \delta^{[\mu}_{[\sigma}
{C_{\rho]}}^{\nu]\kappa\lambda} + \frac{1}{4}
{C_{\mu\nu[\sigma}}^{[\kappa} \delta^{\lambda]}_{\rho]} \label{huit}\CR
\frac{1}{2}{C^{\mu\nu}}_{\theta\tau}\Scal{{{P^-}_{\sigma\rho}}^{\theta\eta}{P^{-\,\tau}}_{\eta\kappa\lambda}}
&=& {{P^-}_{\sigma\rho}}^{[\mu|\eta}{P^{-\,\nu]}}_{\eta\kappa\lambda}
\label{obstru}\CR
\frac{1}{2}{C^{\mu\nu}}_{\theta\tau}\Scal{{{P^-}_{\sigma\rho}}^{\theta\eta}{P^{+\,\tau}}_{\eta\kappa\lambda}}
&=& -3\,{{P^-}_{\sigma\rho}}^{[\mu|\eta}{P^{+\,\nu]}}_{\eta\kappa\lambda}\CR
\eea
Actually, the two last equations state that if we consider 2-form as
Lie algebra elements, the commutator of two antiselfduals ones gives
a selfdual, and the commutator of one antiselfdual and one
selfdual gives an antiselfdual. The first and the fourth can be seen
from seven dimensional gamma matrix point of view, where $\gamma_a$ states
for vector (antiselfdual 2-form) and $\gamma_{ab}$ for $Spin(7)$ Lie
algebra element (selfdual 2-form), we see that both equations
represent the single formula $\gamma_a\gamma_b = \delta_{ab} + 2\gamma_{ab}$.
\section{$\delta_{\mathfrak{k}}$ invariance of the gauge function in $\Omega$
 background}
\label{gf}
In this appendix, we show that the $\delta$ operator (\ref{odelta})
constrains correctly the more general gauge function exactly
renormalizable in four dimensions.
The gauge function contains, a priori, terms involving the two vectors
$\u$ and $\bu$. We will assume that the gauge function does not
depend of the derivatives of 
$\u$ and $\bu$. These vectors have respectively
ghost number $2$ and $-2$. We can decompose the gauge function of
total ghost number 
$-1$ into a sum over the terms of field's ghost number $2i-1$.
\be
\Uppsi = \sum_i \Uppsi_i
\ee
In exactly the same way $\deltac$ decomposes into $\delta_{-1} +
\delta_0 + \delta_1$ and the equation $\deltac \Uppsi = 0$ decomposes
into
\be
\delta_1 \Uppsi_{i-1} + \delta_0 \Uppsi_i + \delta_{-1} \Uppsi_{i+1} =
0 \hspace{5mm} , \forall i
\ee
Since we are interested in the equivariant part of the action, we have not
considered the fields $c, \bar c$ and $b$. The possible gauge functions
of given field's ghost number are 
\bea 
&\Uppsi_{-1} &= \int_M \trace \Scal{ \a_{-1} i_{\u} \chi \star d_A
\bar\Phi + \b_{-1} \star \eta \Lc_{\u} \bar\Phi } \CR
&\Uppsi_0 &= \int_M \trace \biggl( \a_0 \chi \star T + \b_0 \chi \star F
+ \c_0 \Psi \star d_A \bar\Phi + \d_0 \star \eta [\Phi, \bar\Phi] \biggr. \CR
& & \hspace{15mm} + \a_0' i_{\bu} \chi \star i_{\u} T + \bar \a_0' i_{\u}
\chi \star i_{\bu} T + \b_0' i_{\bu} \chi \star
i_{\u} \scal{F+\beta \star C F} \CR 
& & \hspace{25mm} + \bar \b_0' i_{\u} \chi
\star i_{\bu} \scal{F + \bar \beta \star C F} + \c_0' \star i_{\bu} \Psi \Lc_{\u} \bar\Phi + \bar \c_0' \star
i_{\u} \Psi \Lc_{\bu} \bar\Phi \CR 
& & \hspace{35mm} + \d_0' \star \eta
i_{\bu} i_{\u} T + \e_0' \star \eta i_{\bu} i_{\u} \scal{F+ \gamma
 \star C F} + \f_0' \star i_{\bu} i_{\u} \chi [\Phi, \bar\Phi] \biggr)
\CR
&\Uppsi_1 &= \int_M \trace \Biggl( \a_1 i_{\bu} T \star \Psi + \b_1
i_{\bu} \scal{ F + \alpha \star C F} \star \Psi + \c_1 i_{\bu}
\chi \star d_A \Phi \Biggr .\CR
& & \hspace{65mm} + \d_1 \star \eta \Lc_{\bu} \Phi + \e_1 \star
i_{\bu} \Psi [\Phi, \bar\Phi] \CR
& & \hspace{25mm} + |\bu|^2 \biggl(\a_1' i_{\u} T \star \Psi + \b_1'
i_{\u} \scal{ F + \alpha' \star C F} \star \Psi \biggr . \CR 
& & \hspace{45mm} + \c_1' i_{\u}
\chi \star d_A \Phi + \d_1' \star \eta \Lc_{\u} \Phi + \e_1' \star
i_{\u} \Psi [\Phi, \bar\Phi]\biggr) \Biggr ) \CR
&\Uppsi_2 &= \int_M \trace \Scal{ \a_2 |\bu|^2 \Psi \star d_A \Phi +
 \b_2 \star i_{\bu} \Psi \Lc_{\bu} \Phi + |\bu|^2 \star \scal{ \c_2
 i_{\bu} \Psi \Lc_{\u} \Phi + \bar \c_2 i_{\u} \Psi \Lc_{\bu} \Phi
 }} \CR
\eea 
where the parameters could be arbitrary functions of $(\u\cdot \bu)$. In
this computation, we use the fact that 
$\u$ and $\bu$ are two commuting Killing vectors, the fact that
$\kappa$ is covariantly constant and, as a matter of fact, we must add
to these requirements that these three vectors are linearly
independent, that the three $2$-form $g(\u)_{\, \wedge} g(\bu)$,
$g(\kappa)_{\, \wedge} g(\u)$ and $g(\kappa)_{\, \wedge} g(\bu)$, are
not selfdual, as well as the condition that $(\kappa\cdot \u)$ is not zero.
The easiest way to compute $\deltac \Uppsi = 0$ is to begin by the term
of higher degree in order to constrain the parameters before to
compute the more complex expressions. $\delta_{-1} \Uppsi_{-1} = 0$ does
not give any informations, but $\delta_1 \Uppsi_2 = 0$ constrains
$\a_2$ to be zero. $\delta_1 \Uppsi_1 + \delta_0 \Uppsi_2 = 0$ then
establishes that $\Uppsi_2$ is null, and that $\b_1$ and $\d_1$ are the
only two non zero parameters of $\Uppsi_1$ and must be opposed. Next
$\delta_0 \Uppsi_{-1} + \delta_{-1} \Uppsi_0 = 0$ gives $\a_{-1}$,
$\bar\c_0 '$ and $\f_0 '$ to be zero and constrains $\b_{-1}$, $\c_0$
and $\d_0$ to be equal. The most meaningful equation is $\delta_1
\Uppsi_0 + \delta_1 \Uppsi_0 = 0$ , it gives $\a_0' = \bar
\a_0' = \b_0' = \bar \b_0' = \c_0' = \d_0 = 0$ and constrains all the
other parameters. But, let us look at a residual term more
closely. After all the coefficients have been 
constrained up to a global factor we leave with the expression
\be \delta_1 \Uppsi_0 + \delta_0 \Uppsi_1 = \int_M \frac{\b_0}{2}
g(\kappa)_{\, \wedge} C_{\, \wedge} i_{\bu} \trace \scal{ F_{\,
 \wedge} F} \label{cdp}\ee
which seems to be non zero at first sight. Since the manifold on which
the theory is defined admits a covariantly constant vector field, it
must decompose into $ \mathds{R} \times N$ in the simply
connected case\footnote{See page \pageref{reduced}.}, as
a matter of fact we will be able to annul the term (\ref{cdp}) only in this
case. The embeddings of $\mathds{R}$ in $M$ can be put into effect by
the flow of $\kappa$, which we remind to be defined by
\be \frac{d\,}{dt} \, \phi_{\kappa, t}(p) = \kappa_{|\phi_{\kappa, t}(p)} \ee
This one defines a global function $t$ on $M$, for which the flow
equation can be written $d t = g(\kappa)$. Therefore $g(\kappa)$ is
$d$-exact and we can use this property to compute that\footnote{Note
 that in order to the integration by part to work, $F$ must converges
 as quickly as necessary where $t \rightarrow \infty$ in such way
 that the $t$ factor could be compensated.} 
\be 
\int_M g(\kappa)_{\, \wedge} C_{\, \wedge} i_{\bu} \trace \scal{
 F_{\, \wedge} F} = \int_M \scal{ \, (\kappa\cdot \bu) C
 + t \L_{\bu} C }_{\, \wedge}\trace \scal{
 F_{\, \wedge} F} \ee
In order to this term to be null, $(\kappa\cdot \bu)$ must be constant for
a gauge fiber bundle of trivial second Chern class, and zero
otherwise, and $d g(\bu)$ must be selfdual in order for $\L_{\bu}$ to
leave $C$ invariant. That is why we have not assumed that
$(\kappa\cdot \bu)$ was non zero in the calculus. These requirements are
a bit strong , because they impose $\bu$ to be zero in four
dimensions. Nevertheless the case studied
of Nekrasov 
\cite{nical,nical2} fall in this restricted class. As a matter of fact $\bu$
does not contribute to topological amplitudes by construction. To finish the
computation, note that the last constraint, confirms what has already
been given by the others. Therefore we obtain the well constrained
gauge function up to a global scale factor \begin{multline}
\hspace{1cm} \Uppsi = \int_M \trace \Biggl( \frac{2}{n} \chi\star T +
\chi \star F 
+ \Psi \star \Scal{d_A \bar\Phi - i_{\bu} F} \Biggr . \\*
+ \star \eta \Scal{ [\Phi, \bar\Phi] + \Lc_{\u} \bar\Phi - \Lc_{\bu}
 \Phi + i_{\bu}
 i_{\u} F } \Biggr) \hspace{1cm} 
\end{multline}
and we have obtained the $\delta_{\mathfrak{k}}$-exact gauge function
(\ref{gaom}). 

\section{$\Omega$ background in Euclidean space}
In the case of a flat space, we can write explicitly the form of the
Killing vectors as generators of $\mathfrak{so}(n)$ elements. We give
here the explicit form of the supersymmetric action in the eight
dimensional case. We use the scalar $\Phi$ and $\bar\Phi$ used in the
twisted version, as well as the matrix $\Omega_{\mu\nu} \equiv
\frac{1}{2} \scal{d g(\u)}_{\mu\nu}$ and $\overline{\Omega}_{\mu\nu} \equiv
\frac{1}{2} \scal{d g(\bu)}_{\mu\nu}$. With these definitions the
$\Omega$ background action can be expended as follows :
\begin{multline}
S = \, S_0 \, + \, {\overline{\Omega}^\mu}_\nu {\overline{S}_{1\,
 \mu}}^\nu \, + \, 
 {\Omega^\mu}_\nu {S_{1\, \mu}}^\nu \\* + {\Omega^\mu}_\sigma
{\Omega^\nu}_\rho {S_{2\, \mu\nu}}^{\sigma\rho} + {\Omega^\mu}_\sigma
{\overline{\Omega}^\nu}_\rho {S^{\scriptscriptstyle (1,1)}_{2\, \mu\nu}}^{\sigma\rho} + {\overline{\Omega}^\mu}_\sigma
{\overline{\Omega}^\nu}_\rho {\overline{S}_{2\, \mu\nu}}^{\sigma\rho} \\*+
{\overline{\Omega}^\mu}_\sigma {\overline{\Omega}^\nu}_\rho
{\Omega^\kappa}_\lambda {\overline{S}_{3\, \mu\nu\kappa}}^{\sigma\rho\lambda} +
{\Omega^\mu}_\sigma {\Omega^\nu}_\rho {\overline{\Omega}^\kappa}_\lambda
{S_{3\, \mu\nu\kappa}}^{\sigma\rho\lambda} \\*+
{\Omega^\mu}_\sigma {\overline{\Omega}^\nu}_\rho {\Omega^\kappa}_\theta
{\overline{\Omega}^\lambda}_\tau {S_{4\,
 \mu\nu\kappa\lambda}}^{\sigma\rho\theta\tau} 
\end{multline} 
with the definitions 
\bea
&S_0 &\equiv \int_M d^8x \trace \biggl( -\frac{1}{4} F_{\mu\nu} F^{\mu\nu} +
 D_\mu \bar\Phi D^\mu \Phi - \frac{i}{2} \lambda^\alpha \baa
 D_{\alpha\aalpha} \lambda^{\aalpha} - \frac{i}{2} \lambda_{\aalpha}
 \baa D^{\aalpha\alpha} \lambda_{\alpha} \biggr . \CR
& & \hspace{6cm} + \frac{1}{2} [\Phi, \bar\Phi]^2 + \lambda_{\aalpha}
[\bar\Phi, 
\lambda^{\aalpha}] - \frac{1}{2} \lambda^\alpha [\Phi, \lambda_\alpha]
\biggr) \CR
&{\overline{S}_{1\, \mu}}^\nu &\equiv \int_M d^8x \biggl( \, x^\nu
\trace \scal{ F_{\mu\sigma} 
 D^\sigma \Phi + [\Phi, \bar\Phi] D_\mu \Phi - \lambda_{\aalpha} D_\mu
 \lambda^{\aalpha}} + \frac{1}{2} \trace \scal{ \lambda_{\aalpha}
 {{\gamma_\mu}^{\nu\aalpha}}_{\bbeta} \lambda^{\bbeta}} \biggr) \CR
&{S_{1\, \mu}}^\nu &\equiv \int_M d^8x \biggl( \, x^\nu \trace \scal{
 F_{\mu\sigma} 
 D^\sigma \bar\Phi - [\Phi, \bar\Phi] D_\mu \bar\Phi + \frac{1}{2}
 \lambda^\alpha D_\mu 
 \lambda_\alpha} - \frac{1}{4} \trace \scal{ \lambda^\alpha
 {{{\gamma_\mu}^\nu}_\alpha}^\beta \lambda_\beta} \biggr) \CR
&{S_{2\, \mu\nu}}^{\sigma\rho} &\equiv \frac{1}{2} \int_M d^8 x\, \,
x^\sigma x^\rho \, \trace \scal{ D_\mu \bar\Phi D_\nu \bar\Phi } \CR
&{S^{\scriptscriptstyle (1,1)}_{2\, \mu\nu}}^{\sigma\rho} &\equiv
\int_M d^8x\, \, x^\sigma x^\rho \, \trace 
\scal{ {F_\mu}^\kappa F_{\nu\kappa} + F_{\mu\nu} [\Phi, \bar\Phi] -
 D_\mu \bar\Phi 
 D_\nu \Phi} \CR
&{\overline{S}_{2\, \mu\nu}}^{\sigma\rho} &\equiv \frac{1}{2} \int_M
d^8 x\, \, x^\sigma x^\rho \, \trace \scal{ D_\mu \Phi D_\nu \Phi } \CR
&{\overline{S}_{3\, \mu\nu\kappa}}^{\sigma\rho\lambda} &\equiv \int_M d^8x
\, \, x^\sigma x^\rho x^\lambda \,\trace \scal{ F_{\kappa\mu} D_\nu \Phi} \CR
&{S_{3\, \mu\nu\kappa}}^{\sigma\rho\lambda} &\equiv \int_M d^8x \, \, 
x^\sigma x^\rho x^\lambda \, \trace \scal{ F_{\kappa\mu} D_\nu \bar\Phi} \CR
&{S_{4\, \mu\nu\kappa\lambda}}^{\sigma\rho\theta\tau} &\equiv
\frac{1}{2} \int_M d^8x \, \, 
x^\sigma x^\rho x^\theta x^\tau \, \trace \scal{ F_{\mu\nu} F_{\kappa\lambda}}
\CR
\eea
\subsection*{Acknowledgments}

This work was partially supported under the contract ANR(CNRS-USAR) \\ \texttt{no.05-BLAN-0079-01}.

\end{document}